\documentclass{elsart}

\usepackage{epsfig}
\usepackage{subfigure}
\usepackage{amssymb}

\newcommand{\ns}{\ensuremath{~\rm{ns}}}

\newcommand{\ms}{\ensuremath{~\rm{ms}}}

\newcommand{\m}{\ensuremath{~\rm{m}}}
\newcommand{\mm}{\ensuremath{~\rm{mm}}}
\newcommand{\um}{\ensuremath{~\rm{\mu m}}}
\newcommand{\nm}{\ensuremath{~\rm{nm}}}

\newcommand{\GeV}{\ensuremath{~\rm{GeV}}}
\newcommand{\MeV}{\ensuremath{~\rm{MeV}}}

\newcommand{\GHz}{\ensuremath{~\rm{GHz}}}
\newcommand{\MHz}{\ensuremath{~\rm{MHz}}}
\newcommand{\kHz}{\ensuremath{~\rm{kHz}}}
\newcommand{\Hz}{\ensuremath{~\rm{Hz}}}

\newcommand{\dB}{\ensuremath{~\rm{dB}}}

\newcommand{\E}[1]{\ensuremath{\times 10^{#1}}}

\newcommand{\mrad}{\ensuremath{~\rm{mrad}}}
\newcommand{\urad}{\ensuremath{~\mu \rm{rad}}}

\newcommand{\degC}{\ensuremath{^{\circ} {\rm C}}}

\begin{document}

\begin{frontmatter}

\title{Cavity BPM System Tests for the ILC Energy Spectrometer}
\author[add:cambridge]{M.~Slater\corauthref{corr_info}}
 \ead{slater@hep.phy.cam.ac.uk}
\author[add:slac]{C.~Adolphsen}
\author[add:slac]{R.~Arnold}
\author[add:rhul]{S.~Boogert}
\author[add:rhul]{G.~Boorman}
\author[add:ucl]{F.~Gournaris}
\author[add:nd]{M.~Hildreth}
\author[add:lbnl]{C.~Hlaing}
\author[add:daresbury]{F.~Jackson}
\author[add:lbnl]{O.~Khainovski}
\author[add:lbnl]{Yu.~G.~Kolomensky}
\author[add:ucl]{A.~Lyapin}
\author[add:ucl]{B.~Maiheu}
\author[add:slac]{D.~McCormick}
\author[add:ucl]{D.~J.~Miller}
\author[add:lbnl,add:caltech]{T.~J.~Orimoto}
\author[add:slac]{Z.~Szalata}
\author[add:cambridge]{M.~Thomson}
\author[add:cambridge]{D.~Ward}
\author[add:ucl]{M.~Wing}
\author[add:slac]{M.~Woods}

\address[add:lbnl]{University of California and Lawrence Berkeley National Laboratory,
Berkeley, California, USA\thanksref{fund:lbnl}}
\address[add:caltech]{California Institute of Technology, Pasadena, California, USA}
\address[add:cambridge]{University of Cambridge, Cambridge, UK\thanksref{fund:uk}}
\address[add:daresbury]{Daresbury Laboratory, Daresbury, UK\thanksref{fund:uk}}
\address[add:ucl]{University College London, London, UK\thanksref{fund:uk}}
\address[add:nd]{University of Notre Dame, Notre Dame, Indiana, USA}
\address[add:rhul]{Royal Holloway, University of London, Egham, UK\thanksref{fund:uk}}
\address[add:slac]{Stanford Linear Accelerator Centre, Menlo Park, California, USA\thanksref{fund:slac}}
\corauth[corr_info]{Corresponding Author. HEP Group, University of Cambridge, 
Cambridge, UK, CB3 0HE. Tel:
+44 (0)1223 337475; Fax: +44 (0)1223 353920}

\thanks[fund:uk]{This work was supported by the Commission of the European Communities under the 6th Framework Programme ``Structuring the European Research Arm,'' contract number RIDS-011899 and by the Science and Technology Facilities Council (STFC)}
\thanks[fund:slac]{This work was supported by the U.S. Department of Energy under contract DE-AC02-76SF00515}
\thanks[fund:lbnl]{This work was supported by the U.S. Department of Energy under contract DE-FG02-03ER41279}

\begin{abstract}
The main physics programme of the International Linear Collider (ILC) requires a 
measurement of the beam energy at the interaction point with an accuracy of $10^{-4}$ or 
better. To achieve this goal a magnetic spectrometer using high resolution beam position 
monitors (BPMs) has been proposed. This paper reports on the cavity BPM system that was 
deployed to test this proposal. We demonstrate sub-micron resolution and micron level 
stability over 20 hours for a $1\m$ long BPM triplet. We find micron-level stability over
1 hour for 3 BPM stations distributed over a $30\m$ long baseline.
The understanding of the behaviour and response of the BPMs gained from this 
work has allowed full spectrometer tests to be carried out.
\end{abstract}


\end{frontmatter}

\section{Introduction}

The physics potential of the International Linear Collider (ILC) depends greatly on 
precise energy measurements of the electron and positron beams at the interaction 
point (IP). Two types of analysis are particularly sensitive to the collision energy: 
threshold cross-section measurement and reconstruction of particle resonances 
\cite{ref:theory_measurements}. The required accuracy for the mass measurements dictate 
that the fractional error on the determination of the incoming beam energy must be better than 
$10^{-4}$. To measure the energy to this level and to minimise the disruption 
of the beam, a magnetic spectrometer has been proposed.

When passing through a magnetic field, a particle with an electric charge $q$ is 
deflected by an angle $\theta$ which is inversely proportional to its energy $E$:
\begin{equation}
\label{eq:spec_eq}
\theta = \frac{c~q}{E} \int{{\it B} \cdot {\rm d} \ell}
\end{equation}
\noindent where $c$ is the speed of light, ${\it B}$ is the magnetic field and 
${\rm d} \ell$ is the path segment along which the particle travels. The initial 
ILC spectrometer proposal~\cite{ref:lcdet04-31} suggested using a single bend and,
with careful characterisation of the bending magnets and an accurate measurement of
the bend angle $\theta$, the energy of the beam could then be reconstructed.

The performance of a similar spectrometer has already been demonstrated
during the second phase of the Large Electron Positron (LEP2) collider at 
CERN \cite{ref:lepnote}. The magnetic spectrometer installed at LEP2
achieved an accuracy of $1.7\E{-4}$ by measuring the
change in bend angle as the beam passed through a single steel dipole
magnet. To obtain the high accuracy required, a relative energy
measurement was made with the spectrometer calibrated at the $Z^0$
resonance using the resonant depolarisation method, thus
removing the need for an absolute angle measurement. In order to
further improve the accuracy of the measurement, an average was taken
over many bunches and many revolutions of the beam around the
accelerator. This allowed the requirements on the BPM resolution to be
relatively loose. 

\begin{figure}
\begin{center}
\epsfig{file=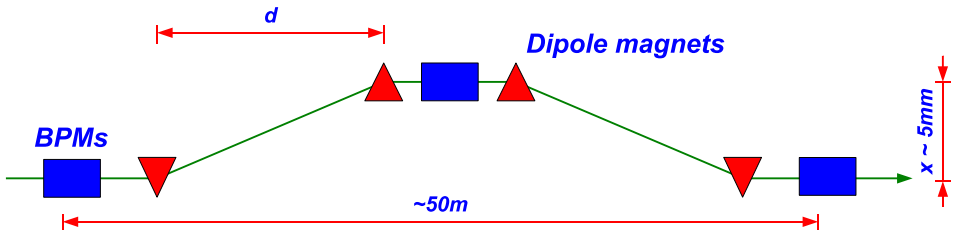,width=\textwidth}
\caption{\label{fig:spectrometer} Schematic of the current baseline design for
the ILC spectrometer.}
\end{center}
\end{figure}

As the ILC is a linear machine, resonant depolarisation is not an
option for the beam energy measurement and in order to limit low
energy operation, the ILC spectrometer will have to provide an
absolute energy measurement. In addition, a bunch-by-bunch energy
measurement is highly desirable to remove some of the dependence on
the stability of the machine during physics data taking. These
constraints, in addition to the required resolution of the energy
measurement ($10^{-4}$), force the performance of the combined
BPM and electronics to be considerably higher than the system used at LEP2.
In order to achieve this level of performance, high resolution cavity 
BPMs will need to be used. However, as cavity BPMs are sensitive to the beam slope as 
well as the beam offset, we are focusing our studies on a different approach to the
energy measurement problem from the single bend method initially
proposed. Two identical magnets with opposite fields allow the beam
energy to be measured as a function of the induced horizontal
displacement $x$ given by
\begin{equation}
E = \frac{c~q~d}{x} \int{{\it B}~ {\rm d} \ell}.
\end{equation}
\noindent where $d$ is the total bend length and the deflection angle $\theta$ is small
(fig.~\ref{fig:spectrometer}). The inclination of the beam trajectory through the BPMs 
is then minimised and, by mounting the mid-chicane 
BPMs on high precision movers, they could be translated to the beam location 
in the case of large induced offsets.
The precision and accuracy of the measured offset $x$ contributes directly to the 
uncertainty of the final energy measurement. The current baseline spectrometer 
design~\cite{ref:baseline} implements a $5\mm$ deflection. An offset 
measurement to better than $500\nm$ is therefore required to achieve the necessary energy 
resolution. However, a BPM system with better performance would allow a smaller 
deflection and, therefore, reduce the emittance growth due to 
synchrotron radiation. In addition, the system has to be stable to the level of $500\nm$ over 
multiple hours to avoid extensive recalibration and consequent loss of luminosity.

The operation and stability of a long baseline BPM system in the proposed spectrometer is of 
interest in several other fields. For example, the beam position in the ILC linacs must be 
measured with a resolution of $1\um$ for the orbit correction and emittance preservation. High 
resolution BPMs are also required throughout the beam delivery system (BDS). Current test 
facilities and modern linac based light sources are also demanding
significantly better performance from their BPM systems. Most notably,
the ATF2 \cite{ref:atf2} facility at KEK, designed to test the final
focus optics for the ILC, not only requires resolutions of
$\sim100\nm$ for the extraction line BPMs but good stability and ease
of use of the entire monitoring system. 

In the framework of the T-474 test beam experiment \cite{ref:proposal} in End Station A (ESA) 
at the Stanford Linear Accelerator Center (SLAC), we installed several BPM stations in 
the $40\m$ available drift space. The aim of three running periods in
2006 was to commission and optimise these BPMs and study their
resolution and stability. The principal results of these runs are
discussed below.

\section{ESA beam and hardware configuration}
\label{sect:beamline_setup}

\subsection{Beam delivery to End Station A}

\begin{figure}
\begin{center}
\epsfig{file=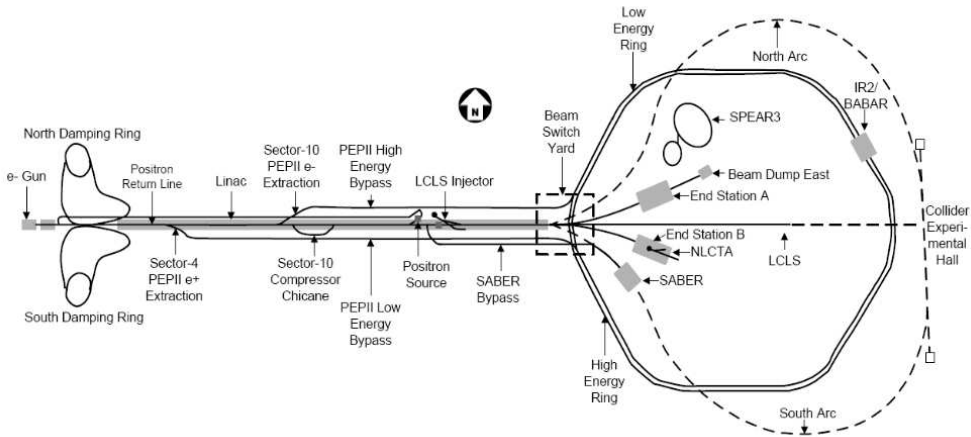,width=\textwidth}
\caption{\label{fig:slac_layout} Schematic of the SLAC beam lines that provide a high energy 
electron beam from the 2-mile SLAC linac, through the Beam Switch Yard to the A-line 
leading to End Station A.}
\end{center}
\end{figure}

SLAC's accelerator and beam transport system for delivering high energy electron beams 
are depicted in Figure~\ref{fig:slac_layout}. A high intensity electron 
beam is produced by a thermionic gun, bunched and pre-accelerated in
the first sections of the linac to $1.19\GeV$. The bunches are then
kicked by a pulsed magnet into the Linac-to-Ring transfer line and then
transported to the electron North Damping Ring (DR), where they are
stored for $8\ms$ to reduce the beam emittance. The Ring-to-Linac
transfer line transports the bunches back from the DR and a pulsed
magnet kicks them into the linac at Sector 2. The beam is subsequently accelerated to 
$28.5\GeV$. At the end of the linac it is then transported from the
Beam Switch Yard (BSY) through a $24.5^{\circ}$ bending section into
the A-line on its way to End Station A. ESA test beams operate with 
single bunches at $10\Hz$ parasitically to PEP-II operation. The ESA
is currently the highest energy test beam facility available, with its
other beam parameters, listed in Table~\ref{table:beam_comparison}, similar to
those for the ILC. 

\begin{table}
\begin{center} 
\caption{\label{table:beam_comparison} Beam parameters at ESA and
as proposed for the ILC~\cite{ref:mikes_proposal}.}
\vspace*{0.1cm}
\begin{tabular}{c|c|c} \hline
Parameter & SLAC ESA & ILC-500 \\ \hline \hline
Repetition rate & $10\Hz$ & $5\Hz$ \\
Energy & $28.5\GeV$ & $250 \GeV$ \\
Train length & up to $400\ns$ & $1 \ms$\\
Micro bunch spacing & $20-400\ns$ & $337\ns$ \\
Bunches per train & 1 or 2 & 2820\\
Bunch charge & $1.6\E{10}$ & $2.0\E{10}$\\
Bunch length & $500\um$ & $300\um$\\
Energy spread & $0.15\%$ & $0.1\%$ \\
\hline
\end{tabular}
\end{center}
\end{table}

\begin{figure}
\begin{center}
\epsfig{file=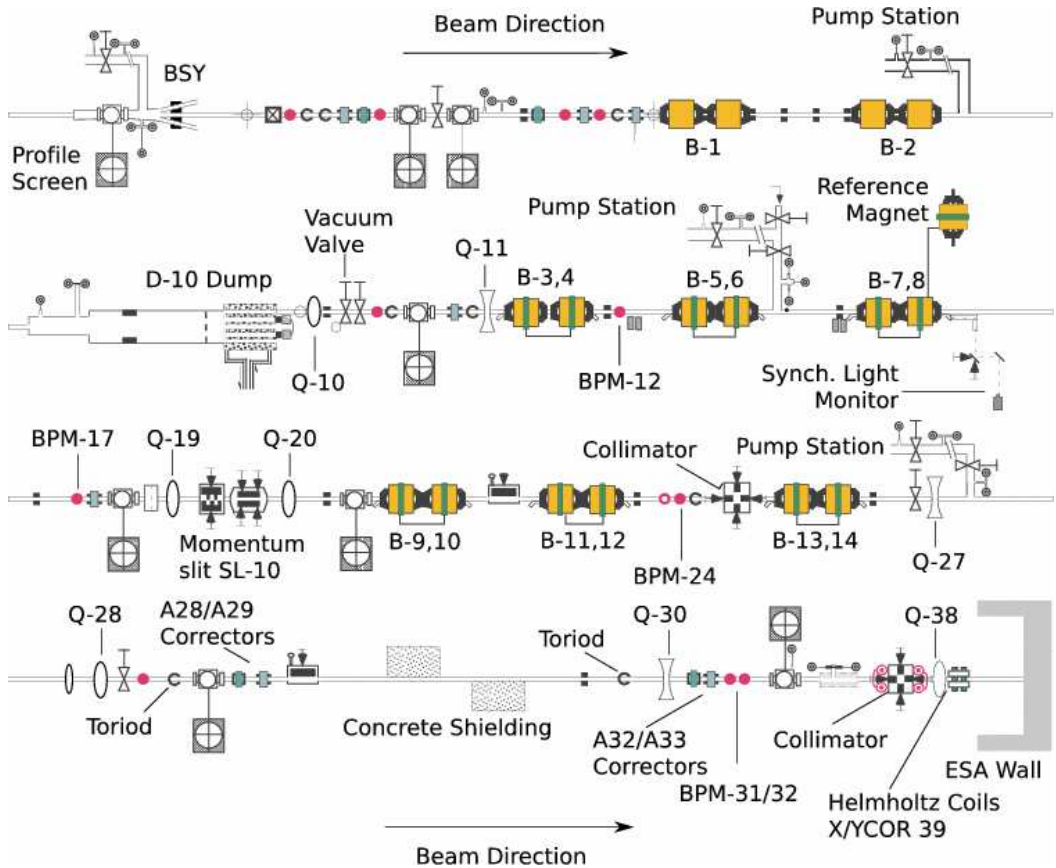,width=\textwidth}
\caption{\label{fig:aline_config} The beam line configuration for the
  A-line. The beam goes from left to right, top to bottom.}
\end{center}
\end{figure}

\begin{figure}
\begin{center}
\vspace*{-0.5cm}
\epsfig{file=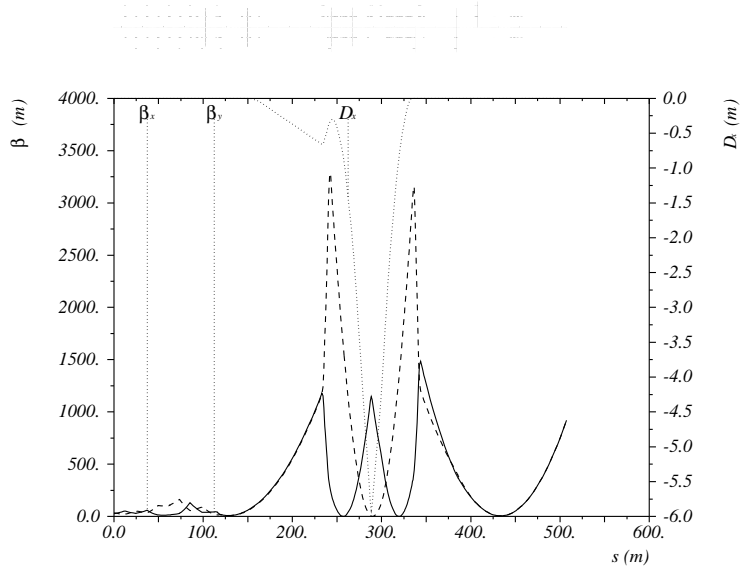,width=0.8\textwidth}
\vspace*{-1.5cm}
\caption{\label{fig:beta_dispersion} The beta functions ($\beta_x$, $\beta_y$)
and horizontal dispersion ($D_x$) in the A-line and ESA. The origin is
the swivel point in the BSY. The maximum dispersion is at the SL-19
momentum slits in the Aline, as well as the horizontal waist.}
\end{center}
\end{figure}

The A-line leading from the BSY to ESA is about $300\m$ long (see
Figure~\ref{fig:aline_config}). Two BSY dipoles, B1 and B2,  just upstream of the
D-10 beam dump at the start of the A-line bend the beam through an
initial angle of $0.5^{\circ}$, and are followed by a string of 12
dipoles which deflect the beam further by $24^{\circ}$.  
The quadrupole lattice in the A-line consists of a doublet, Q10 and Q11, at the start of the 
A-line which brings the beam to a horizontal waist at the dispersion
matching quadrupoles, Q19 and Q20. These are located at the high
dispersion point (see Figure~\ref{fig:beta_dispersion}) in the middle of the
A-line to correct first- and second-order horizontal dispersion in ESA.
Four additional quadrupoles (Q27, Q28, Q30, Q38) near the end of the
A-line control the transversal beam size in ESA. The beta functions and horizontal
dispersion through the A-line and ESA are shown in
Figure~\ref{fig:beta_dispersion}. Vertical and horizontal corrector
dipole pairs, A28/A29 and A32/A33 are used to set the beam trajectory
in ESA. They are used in the SLAC Control Program (SCP) feedback to
stabilise the beam position at BPM 31 at the end of the A-line and BPM
1 in ESA. SL-10 is a high power momentum slit located at the A-line
bend mid-point where the dispersion is at maximum. The maximum momentum
acceptance is $2\%$. A SCP energy feedback system uses a stripline
BPM, BPM17, at a high dispersion point near SL-10 to stabilise the
beam energy.

A-line beam diagnostics include RF cavity BPMs, charge-sensitive
toroids, a synchrotron light monitor, retractable profile screens, and
high frequency diodes.  The synchrotron light monitor is positioned
upstream of the mid-bend region, and is used to monitor beam energy
spread and provide a diagnostic for minimising it. High frequency
diodes are installed in ESA for monitoring and tuning the bunch
length~\cite{ref:slac1}.

\subsection{ESA beam line and experimental equipment}

\begin{figure}
\begin{center}
\epsfig{file=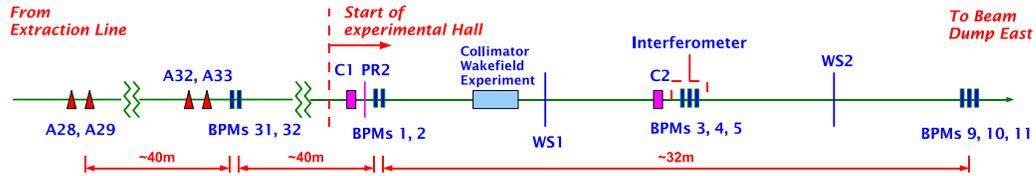,width=\textwidth}
\caption{\label{fig:beamline} Schematic of the principal beam line components at the end of 
  the A-line and in ESA.}
\end{center}
\end{figure}

The configuration of beam line components, beam diagnostics and experimental equipment at 
the end of the A-line and in ESA is shown in Figure~\ref{fig:beamline}. Two protection 
collimators are located in ESA; 3C1 ($19\mm$ aperture radius) is at the entrance to ESA in front of 
BPM 1 and 3C2 ($8\mm$ aperture radius) is located in front of BPM 3.  There are two beam 
profile monitors, one upstream of the T480 collimator wakefield
experiment (PR2) and one just beyond the east wall of ESA (PR4). These
are aluminium oxide screens that can be inserted into the beampipe.  Two
wire scanners, WS1 and WS2, are approximately $20\m$ apart and are
used to perform transverse beam size measurements. $75\um$ tungsten wires
are scanned across the beam, creating low energy electrons that are
detected by photomultiplier tubes at $90^{\circ}$.  Several beam loss
monitors along the beam line in ESA are interlocked to the machine and
radiation protection systems. 

Our experimental equipment includes a BPM doublet (BPMs 1,2) and two BPM triplets 
(BPMs 3-5 and 9-11) and an interferometer system monitoring horizontal motion of the 
BPM 3-5 triplet.  These systems are described in more detail below. 
Among other experimental setups,  a collimator wakefield experiment is
located downstream of the BPM 1 and 2 doublet and uses the BPM
system for measuring wakefield kick angles~\cite{ref:t480}. 

We measured the beam jitter at various BPM stations along the
beam line. The typical values for the data discussed in this paper are
detailed in Table~\ref{table:jitter}.

\begin{table}
\begin{center}
\caption{\label{table:jitter} The typical beam jitter as recorded at
  the BPM stations for the data considered in this paper. 
  The error indicates the variation seen over a one hour run.} 
\vspace*{0.1cm}
\begin{tabular}{c|c|c} \hline 
BPMs & Jitter $x$ ($\um$) & Jitter $y$ ($\um$) \\ \hline \hline 
1-2  & $110\pm17$ & $14\pm11$ \\
3-5  & $157\pm15$ & $40\pm19$ \\
9-11 & $173\pm16$ & $66\pm22$ \\
\hline
\end{tabular}
\end{center} 
\end{table}

\subsection{Cavity BPMs}
\label{sect:cavity_specs}

A microwave or Radio Frequency (RF) cavity BPM is a discontinuity in the beam line
which, through the excitation of different oscillating electromagnetic field configurations
(cavity modes), can be used to measure the position of the transiting bunch.
It can be shown that the amplitude of the excited dipole field, 
providing the bunch is not far from the centre of the BPM, is not only proportional to the 
charge, but also related to the beam offset $x$ from the centre, the
slope $\theta$ of the beam trajectory and the tilt $\alpha$ of the
bunch with respect to the cavity \cite{ref:yury_whittum,ref:nanobpm}:

\begin{eqnarray}
\nonumber V_x \propto x e^{- \Gamma t} \sin{\omega t}             \\
          V_\alpha \propto - \alpha e^{- \Gamma t} \cos{\omega t} \\
\nonumber V_\theta \propto \theta e^{- \Gamma t} \cos{\omega t}
\end{eqnarray}

\noindent where the $V$s are contributions to the induced voltage, and
$\omega$ and $\Gamma$ are the frequency and the decay constant of the
dipole mode. The combined cavity output is given by:

\begin{eqnarray}
V(t)      & = & V_x + V_\alpha + V_\theta\\
\nonumber & = & e^{- \Gamma t} \left[ A_x x \sin (\omega t) +
  (A_\theta \theta + A_\alpha \alpha) \cos (\omega t) \right],
\label{eq:V}
\end{eqnarray}

\noindent Therefore the position component is always in quadrature with the combined 
slope and tilt signal and consequently, the angle and position
components can be separated providing the beam arrival time is
known. For that purpose and also as a bunch charge reference we use
additional smaller cavities operating with the monopole mode at the
same frequency as the dipole mode in the position cavities. Their
signals follow the same processing path as the BPM signals in order to
preserve the phase relation.

We used two types of cavity BPMs in our studies: rectangular and cylindrical, all of which have 
a similar dipole mode frequency. The loaded quality factor $Q_L$,
which defines the signal decay constant $\Gamma$ via
\begin{equation}
  \Gamma = \frac{\omega}{2Q_L}
\end{equation}
is also similar for all the cavities except for 3,~4 and 5. The
general cavity parameters are summarised in
Table~\ref{table:cavity_specs}. 

\begin{table}
\begin{center} 
\caption{\label{table:cavity_specs} Parameters of the various BPMs along the ESA beam line. 
The distances in brackets given for the location are measured relative
to the entrance of the experimental hall.}
\vspace*{0.1cm}
\begin{tabular}{c|c|c|c|c} \hline
BPM number & Location & Resonant freq. & Loaded $Q_L$ & Aperture\\ 
           &          &  (MHz)         &            &  (mm)  \\ \hline \hline 
31, 32     &  A-Line             & $2856$  & $\sim 3000$ & $51$ \\
1, 2       &  ESA ($\sim 10\m$)  & $2856$  & $\sim 3000$ & $51$(1) and $38$(2)\\
3, 4, 5    &  ESA ($\sim 25\m$)  & $2859$  & $\sim 500$  & $36$ \\
9, 10, 11  &  ESA ($\sim 40\m$)  & $2856$  & $\sim 3000$ & $20$ \\
\hline
\end{tabular}
\end{center}
\end{table}

\begin{figure}[!ht]\centering
  \subfigure{\epsfig{file=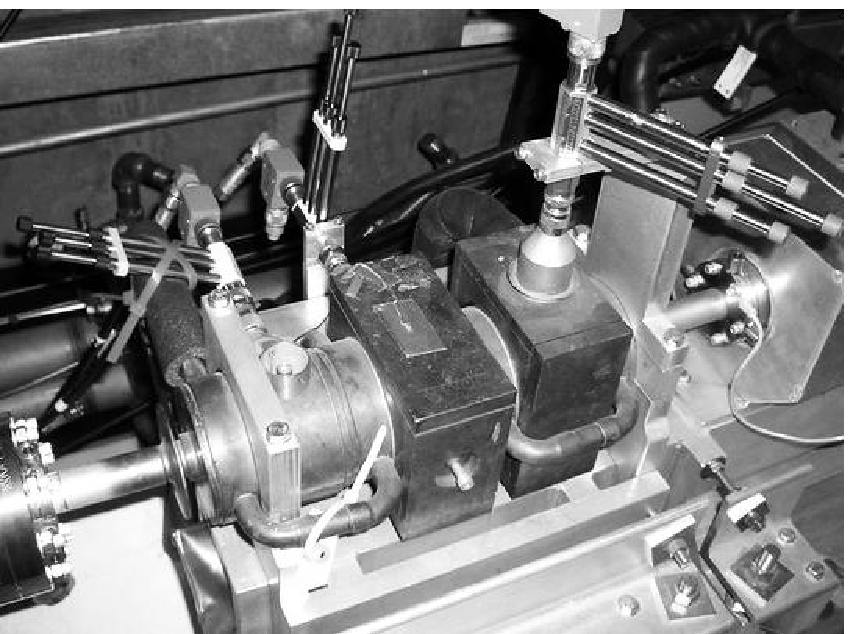,width=0.45\textwidth}}
  \hspace*{0.2cm}
  \subfigure{\epsfig{file=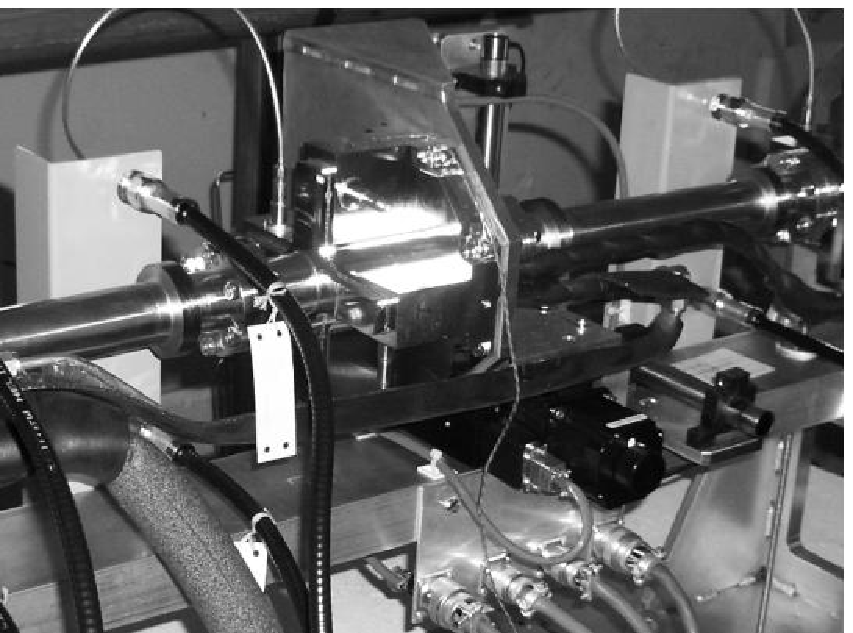,width=0.45\textwidth}}
  \caption{ a) An example of the rectangular cavities used for BPMs 31, 32, 
    1, 2, 9, 10 and 11 showing $x$, $y$ and reference channels and b) the
    central BPM of the prototype ILC linac BPMs triplet. }
\label{fig:slac_cavity}
\end{figure}

Seven rectangular cavity BPMs are available: two in the A-line (BPMs 31 and 32) and five in
the End Station (BPMs 1, 2, 9-11). 
BPMs 9, 10 and 11 were originally designed for use in the 
main SLAC linac whereas BPMs 31, 32, 1 and 2 were built for the A-Line. Each BPM consists 
of three separate cavities (see Figure~\ref{fig:slac_cavity}a): one cylindrical reference 
cavity and two rectangular cavities delivering $x$ and $y$ position dependent signals. 
The rectangular shape of the position cavities helps reduce the
effects of coupling between the $x$ and $y$ orientations. We tuned all
the rectangular BPMs (including their monopole cavities) to a nominal frequency of
$2.856\GHz$ using external tuners. These BPMs were initially installed for the SLAC E158
experiment~\cite{ref:e158}. During this installation, BPM 1 suffered significant
mechanical damage and as a consequence, both the
$x$ and $y$ cavity signals had strong coupling with other modes
resulting in worse resolution and stability.

The central BPM station (BPMs 3-5) consists of three cylindrical cavities designed for use in 
the cryogenic regions of the ILC linac (see 
Figure~\ref{fig:slac_cavity}b)~\cite{ref:ilc_linac_bpms}. In these BPMs, a single cavity is 
used to provide both $x$ and $y$ signals. A combination of slots and waveguide couplers 
provides very good suppression of the unwanted monopole modes. These cavities have a lower 
$Q$-value and therefore a shorter decay time of the dipole mode signal to provide 
bunch-to-bunch position measurements in the ILC without the need to exclude the contamination 
from the previous bunches. We used the reference cavity of BPM 9 to provide the phase and 
charge information for BPMs 3-5; it was therefore tuned to $2.859\GHz$.

\subsection{DAQ and signal processing electronics}
\label{sect:proc_elec}

\begin{figure}
\begin{center}
\epsfig{file=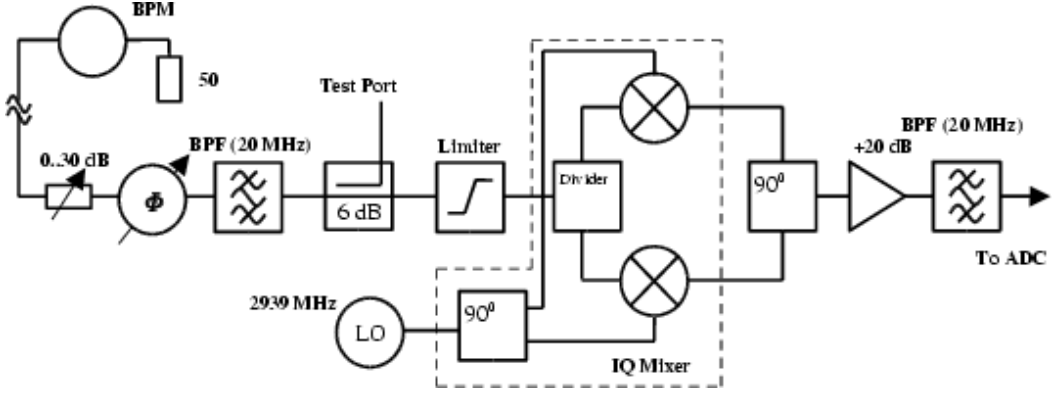,width=\textwidth}
\caption{\label{fig:esa_electronics} Schematic view of the signal processing electronics. The 
components within the dashed lines are contained in a single mixer module}
\end{center}
\end{figure}

The signal processing electronics that convert the $2.9\GHz$ signals coupled out of the 
cavities to low frequency signals prior to digitisation consist of a single stage down-mixing 
circuit (see Figure~\ref{fig:esa_electronics}). The amplitude of the input signal can be 
adjusted manually with variable attenuators and phase adjustment is also possible. The signal 
is filtered at the front end with a narrow $20\MHz$ band-pass filter centred at $2856\MHz$ 
to suppress the noise and unwanted modes. No amplification is applied 
at this stage in order to use the full dynamic range of the mixers. The filtered signal is 
then mixed with a local oscillator (LO) signal operating at $2939\MHz$ to produce a low 
frequency waveform at $\sim83\MHz$. An image reject mixer scheme is used to reduce the 
impact of the unwanted noise and cavity modes around the image frequency. The down-mixed 
signal is then passed through $20\dB$ of amplification and an additional $20\MHz$ bandpass 
filter.

The electronics for BPMs 3-5 and 9-11 are located close to the beam line, about $20\m$ from 
the BPM stations in the experimental hall. The electronics for BPMs 31, 32, 1 and 2 are in a 
different area (``Counting House"), with an additional $\sim30\m$ of cabling required to 
transfer the high frequency cavity signals to the electronics racks. Also, the signals from 
BPMs 31, 1 and 2 are split to provide input for the SLAC Control Program (SCP) running 
the position feedback, adding $6\dB$ attenuation to these channels.

Digitisation is carried out by four VME Struck SIS330x digitisers, two
12 bit (BPMs 31, 32, 1 and 2) and two 14 bit (BPMs 3-5 and 9-11), with
$\pm$2.5 V input range and 50~$\Omega$ input impedance. An external
$119\MHz$ clock signal derived from the linac RF system is used for
all the modules. Consequently, the digitiser under-samples the
signals, producing a waveform aliased to $\sim36\MHz$. Under-sampling
is employed in order to widen the gap between the signal peak and
image sideband, thus simplifying the hardware filtering at the front
end. The digitisers are triggered using a pulse synchronous to the
bunch arrival time, which is generated by the linac control system. 

The data from the SIS digitisers, as well as that from additional gated Analog-to-Digital
Converters (ADCs), VME Smart Analog Monitor (VSAM) and Computer Automated Machine and
Control (CAMAC) modules are captured by a Windows XP PC running Labview DAQ programme
. In addition to the waveform data, some environmental data 
are also available such as information on the linac status from the SCP, temperature data from
thermocouples placed on the BPMs and electronics, interferometer data (see 
Section~\ref{sect:mover}), stray magnetic field data measured with the flux gate probe
and high voltage data for the wire scanners and other experiments. Data are stored at varying 
rates depending on the source: the waveform data are recorded on a per event basis ($10\Hz$) with
each event corresponding to one bunch being delivered from the linac,  
but data provided by the SCP and other additional modules are recorded at a significantly slower 
rate ($\sim0.1\Hz$).

\subsection{Digital signal processing}
\label{sect:signal_proc}

The voltage at the front-end of the analogue electronics given by Equation (\ref{eq:V}) can be
re-written as:

\begin{eqnarray}
  V(t) & =  & A e^{- \Gamma t} \sin (\omega t + \phi)
  \label{eq:basic_wf}
\end{eqnarray}

\noindent where $A = \sqrt{ (A_x x)^2 + (A_\theta \theta + A_\alpha \alpha)^2} $ is the amplitude of the waveform and $\phi = \tan ^{-1} \left( \frac{A_\theta \theta + A_\alpha 
\alpha}{A_x x} \right)$ is the phase.

For simplicity, we can assume that the analogue electronics are ideal and only convert the 
frequency of the waveforms to the first intermediate frequency $\sim 83\MHz$
so that for both the position and reference cavities we have:

\begin{eqnarray}
\label{eq:pos_wf}
V(t) & =  & a e^{- \gamma t} \sin (\omega_{IF1} t + \phi) \\
\label{eq:ref_wf}
V_{\rm ref}(t) & = & a_{\rm ref} e^{- \gamma t} \sin (\omega_{IF1} t + \phi_{\rm ref})
\end{eqnarray}

\noindent where we neglect the differences of the resonant frequencies
and decay constants between the cavities. Note that the decay constant
of the signals is changed by the band pass filters.

The waveforms described by Equations~(\ref{eq:pos_wf}) and~(\ref{eq:ref_wf}) are what appears 
at the front-end of the digitisers. 
Due to under-sampling, the frequency of the waveform experiences 
another conversion to $f_{IF2}=f_s-f_{IF1}$, where $f_s$ is the sampling frequency of the 
digitisers ($119\MHz$ as mentioned above). The amplitude of the resulting waveform remains 
the same.

In order to extract the amplitude and phase information necessary to recover the position, 
this waveform is down-converted again in software by multiplying by a complex LO signal 
at the same frequency as the waveform. This process is called digital down-conversion (DDC). 
It results in a signal describing the envelope of the initial waveform:

\begin{eqnarray}
V_{\rm DDC}(t) & = & a e^{- \gamma t} \sin{(\omega_{IF2} t + \phi)} e^{-i\omega_{IF2} t} \\
\label{eq:ddc_voltage}
\nonumber          & = & \frac{a}{2} e^{- \gamma t} \left[ - e^{i (\phi + \frac{\pi}{2})} + 
                e^{i (2 \omega_{IF2} t + \phi -  \frac{\pi}{2})} \right]
\end{eqnarray}

The mixing process also produces the up-converted component with twice the initial 
frequency. In order to remove this component and further suppress out-of-band noise, 
a digital filter is applied. We chose a Gaussian filter as it is easy
to implement and has a good phase response within the
pass band. Filtering is done in the time domain. As we are interested
in the amplitude of the waveform at only one sample point, we reduce
the computation time by restricting the filtering to one summation out
of the whole convolution:

\begin{equation}
  \label{eq:ddc_filtering}
  V'_{\rm DDC}(t_0) = \sum_{n=-k}^{n=+k} V_{\rm DDC} (t_n) \cdot F(t_n - t_0)
\end{equation}

\noindent where $V'_{\rm DDC}$ is the filtered voltage response after down-conversion at the time 
$t_0$, $k$ defines the window of the filter, outside which the contribution is neglected 
(we set the limit to 0.1\% of the maximum filter response), and the filter function is given by:

\begin{equation}
  \label{eq:filter_function}
  F(t) = \frac{\sqrt{2\pi} \Delta f}{f_s} {\rm exp} \left( \frac{-t^2 (2\pi \Delta f) ^2}{2 f_s^2} \right)
\end{equation}

\noindent where $\Delta f$ is the bandwidth of the filter and $f_s$ is the sampling frequency.

\begin{figure}
\begin{center}
\epsfig{file=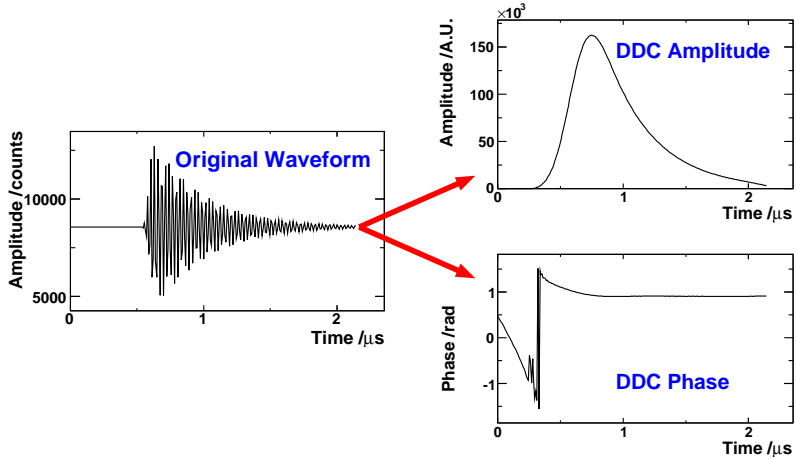,width=0.8\textwidth}
\caption{\label{fig:ddc_pict} A down-converted waveform as recorded at the digitiser as well as 
  the amplitude and phase response of the DDC algorithm for a whole waveform.}
\end{center}
\end{figure}

Applying the filter across all samples we can compute the envelope of the amplitude and the 
phase for the whole waveform (see Figure~\ref{fig:ddc_pict}). A smooth envelope with no 
oscillations and a flat phase response in the region corresponding to large beam power 
are an indication of a good frequency determination and 
low contamination of the waveform with any unwanted signals. The phase variation seen 
prior to this region is a consequence of the broad-band components of the initial beam 
excitation.

The filtered DDC response depends on the amplitude $a$ and phase $\phi$ of the original 
digitised waveform:

\begin{eqnarray}
V'_{\rm DDC}(t_0) \propto a e^{i \phi'} \\
V'_{\rm DDC,ref}(t_0) \propto a_{\rm ref} e^{i \phi'_{\rm ref}}
\end{eqnarray}

\noindent where $\phi'$ includes any phase delays occurring. As mentioned above, the position 
can be extracted with the help of a reference signal. Normalising the complex amplitude from 
the position channel with that from the reference channel, we exclude the charge dependence 
of the position signal:

\begin{equation}
  \label{eq:norm_amp}
  \frac{V'_{\rm DDC}(t_0)}{V'_{DDC,ref}(t_0)}=\frac{a}{a_{\rm ref}} e^{i(\phi'-\phi'_{\rm ref})}=\frac{a}{a_{\rm ref}} e^{i\Delta\phi}
\end{equation}

\noindent where the phase offset $\Delta\phi$ between the position and the reference signals 
at the time $t_0$ does not change unless some changes occur in the processing chain.

In order to recover the position information, we need to calibrate the BPMs applying a known 
offset either to the beam or the BPM. From this we measure the 
scale $S$ (which is the sensitivity of the term $\frac{a}{a_{\rm ref}}$ to the beam offset) and the 
phase $\Phi_{IQ}$ (IQ - in-phase/quadrature rotation) of all the points with a positive offset. 
The position is then given by:

\begin{equation}
  \label{eq:pos_eq}
  x = S~{\rm Re} \left[ \frac{a}{a_{\rm ref}}~e^{i(\Delta\phi - \Phi_{IQ})}\right]
\end{equation}

\noindent Thus, with an accurate calibration the beam position can be found from the amplitude 
and phase of a chosen sample of the waveform. The calibration
coefficients $S$ and $\Phi_{IQ}$ will only stay accurate for a limited 
period of time due to gain and phase drifts in the processing electronics and cables, 
frequency changes of the cavities and other environmental factors. We
therefore took calibration data several times during the running period.

In order to optimise the performance of the BPMs and electronics, both the attenuation
and alignment offsets of the BPMs needed to be minimised. Initially, with the
attenuation set close to maximum to avoid saturation of the waveforms,
we measured the relative misalignment of the BPMs from the nominal
beam trajectory. Using these data we realigned the BPMs, enabling us
to reduce the attenuation. We then repeated this process several times
to optimise the sensitivity and dynamic range. The final offset values
of the BPMs relative to the mean straight line orbit are shown in
Table~\ref{table:alignment}.

\begin{table}
\begin{center}
\caption{\label{table:alignment} The relative misalignment of the
  BPMs with respect to the mean straight line orbit where the error
  quoted is the error on the mean. The alignment of BPM 4 was 
optimised continually through the run as it was mounted on the mover system.}
\vspace*{0.1cm}
\begin{tabular}{c|c|c} \hline
BPM & Offset $x$ ($\mu m$) & Offset $y$ ($\mu m$) \\ \hline \hline 
1 & $-133\pm3$ & $168\pm0$\\
2 & $-90\pm3$ & $-94\pm0$\\
3 & $-128\pm4$ & $-193\pm1$\\
5 & $-134\pm5$ & $10\pm1$\\
9 & $59\pm6$ & $-60\pm2$\\
10 & $125\pm6$ & $29\pm2$\\
11 & $302\pm6$ & $140\pm2$\\
\hline 
\end{tabular}
\end{center} 
\end{table}

\subsection{Mover system and interferometer}
\label{sect:mover}
The middle BPM of the central triplet (BPM 4) is mounted on a dual
axis mover system that allows travel in $x$ and $y$ directions. The
stages each have a $2\mm$ pitch lead screw driven by a 200
steps per revolution stepper motor. This gives a resolution of $10\um$
per step for the BPM moves. The motion system additionally features an
LVDT position readout with an accuracy of $\sim 6\um$. The maximum
allowed travel range of the both stages is limited to about $\pm 6\mm$ for radiation and machine protection reasons. For calibration, a travel distance of $\pm 500\um$ is used, thus giving a
contribution of $\sim1\%$ to the uncertainty on the calibration scale.

\begin{figure}
\begin{center}
\epsfig{file=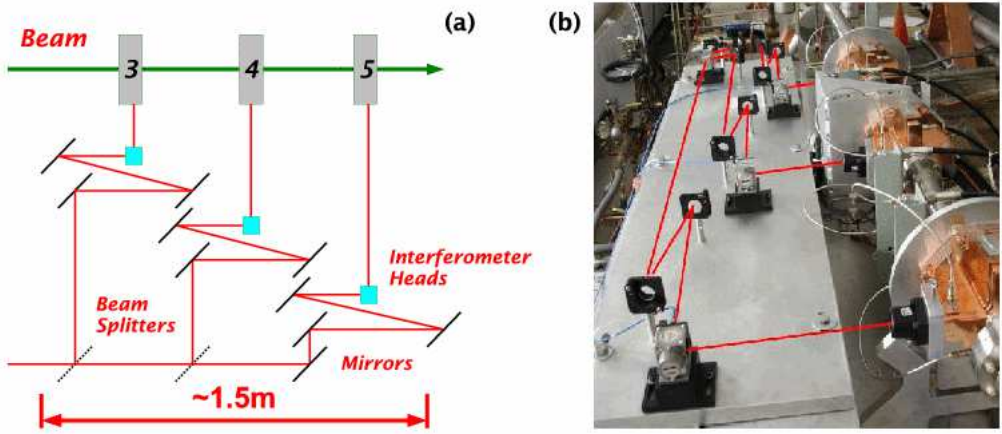,width=\textwidth}
\caption{\label{fig:zygo_layout} The interferometer system in ESA. a) A schematic of the laser path in the interferometer relative to BPMs 3-5 and the beam line. 
  b) The interferometer elements mounted on the optical table with
  BPMs 3-5 on the right. The laser path is indicated schematically.}
\end{center}
\end{figure}

To provide information on the exact $x$ position of BPM 4 and hence
allow corrections to the calibration and mechanical position of the
assembly, we used a set of three single-pass linear 
interferometers~\cite{ref:zygo} manufactured by Zygo. Each
interferometer measures the relative horizontal displacement of the
BPM with respect to interferometer heads on an aluminium table (see
Figure~\ref{fig:zygo_layout}). A single heterodyne helium-neon laser
supplies all three interferometers with polarised light. The
interferometer heads are polarised beam splitters that provide the 
reference and target arms of each of the interferometers with one of
the two laser frequencies. Photo-detectors on a Zygo 4004 VME
measurement board extract the phase and amplitude of the optical 
beat frequency present in the recombined light path, allowing a
measurement of the relative velocity of the BPMs. The measurement
board provides a single bit position resolution of approximately
$0.3\nm$.

Using the interferometer we measured the vibrational motion of BPMs
3-5. This not only includes the rigid motion of the entire system, but
also the non-rigid body motion (i.e. the deviation of each of the
individual BPM positions from a straight line) which contributes
directly to the measured resolution of the BPMs. The amount of motion
recorded at each BPM as well as the non-rigid body motion as extracted
from the interferometer data using an SVD method (described in
Section~\ref{sect:resolution}) are listed in
Table~\ref{table:zygo_motion}. As BPM 4 is mounted on a non-rigid
support, this shows significantly higher vibrational motion than the
other two BPMs.

\begin{table}
\begin{center} 
\caption{\label{table:zygo_motion} Total motion and non-rigid body
  (NRB) motion measured over 30 minutes in the $x$ direction of BPMs
  3-5. The errors quoted are the spread of the RMS. }
\vspace*{0.1cm}
\begin{tabular}{c|c|c} \hline 
BPM & RMS (Total) & RMS (NRB)  \\ \hline \hline 
3 & $170\pm4\nm$ & $94\pm1\nm$ \\
4 & $680\pm37\nm$ & $620\pm10\nm$\\
5 & $130\pm3\nm$ & $72\pm1\nm$\\
\hline
\end{tabular}
\end{center}
\end{table}

\section{Commissioning of the BPMs}

Section \ref{sect:results} will be devoted to the system's performance
during an 18 hour long continuous data taking period which took place at the end of the 2006 running, while in this section we will first report on our
studies of the methodology of dealing with BPM signals. More
specifically, we will discuss the extraction of the calibration
coefficients described in Section~\ref{sect:signal_proc}, namely the
BPM frequency $\omega$, the phase difference between position and
reference signals $\Phi_{IQ}$ and the scale $S$, as well as the optimisation of the DDC algorithm in order to obtain the best resolution and stability of
the BPMs.

For the discussion here and the rest of the analysis results we excluded 
BPM pulses that were saturating the digitisers due to inaccuracies in 
determining the extrapolation factor.

\subsection{BPM frequency and sampling point}
\label{sect:freq_decay}
The frequency of the digitised BPM signal needs to be measured to a
considerable degree of accuracy in order to ensure phase stability
when performing the DDC (see Section~\ref{sect:signal_proc}). 
To determine the optimum frequency for each cavity, we fitted
the waveforms to an exponentially decaying sine wave using
MINUIT~\cite{ref:minuit}. To avoid low amplitude effects, this fit was
only performed on events where the beam was driven to large offsets
($\sim 0.5\mm$, which corresponds to waveforms of approximately half
the dynamic range of the digitiser). We combined the fits from all
such events over an 18 hour period and took the median in order to
get an accurate estimation of the frequency for each BPM throughout
the run. The uncertainties associated with these measurements are
entirely systematic, due to either physical drift and/or systematic
errors associated with the fitting method. We therefore defined the
uncertainty on the frequency as the RMS of its distribution. The
frequencies determined in this way are listed in
Table~\ref{table:freq_full}. 

\begin{table}
\begin{center} 
\caption{\label{table:freq_full} The frequencies that were used throughout the
  18 hour run. The BPMs with errors of zero had a negligible
  systematic error. The frequency for BPM $11x$ is for the first
  8 hours only.}
\vspace*{0.1cm}
\begin{tabular}{c|c|c|c} \hline
BPM & $f_x$ (MHz) & $f_y$ (MHz) & $f_Q$ (MHz) \\ \hline \hline
1 & $36.10\pm0.06$ & $39.47\pm0.03$ & $36.70\pm0.02$ \\
2 & $36.66\pm0.01$ & $36.83\pm0.01$ & as q1 \\
3 & $38.46\pm0.03$ & $37.73\pm0.08$ & $39.16\pm0.01$ \\
4 & $39.38\pm0.07$ & $38.85\pm0.04$ & as q3 \\
5 & $38.94\pm0.15$ & $39.10\pm0.04$ & as q3 \\
9 & $36.53\pm0.01$ & $36.65\pm0.01$ & as q10 \\
10 & $36.65\pm0.01$ & $37.29\pm0.00$ & $37.05\pm0.00$ \\
11 & $36.46\pm0.00$ & $36.84\pm0.01$ & as q10 \\
\hline
\end{tabular}
\end{center} 
\end{table}

\begin{figure}
\begin{center}
\epsfig{file=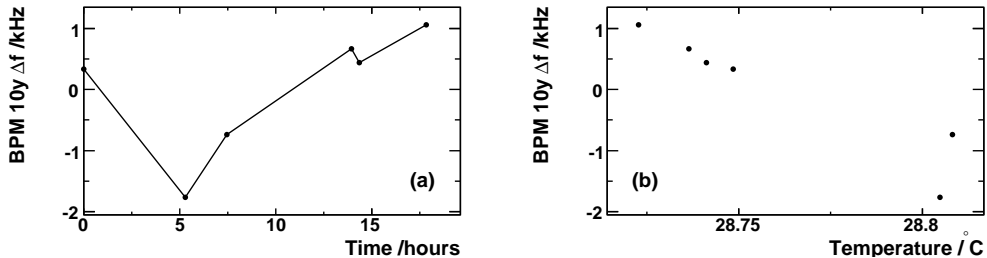,width=1.0\textwidth}
\caption{\label{fig:BPM91011_freq_decay} a) The median frequency 
determined from large offset ($\sim0.5\mm$) events over the 18 hour running period for BPM 
$10y$. b) The fitted frequency versus BPM temperature. }
\end{center}
\end{figure}

The variation in frequencies seen over the 18 hour period was caused
by two different effects. BPMs 3-5 had significantly larger
variation than the majority of the remaining BPMs due to systematic
effects in fitting waveforms with a short decay time. The number of
samples that could be included in the fit was significantly
lower than for the other BPMs, thus leading to a larger error. 
The variation seen in BPMs 9-11 was consistent with being caused by
thermal expansion of the cavities. The frequency of a cavity BPM is
roughly linearly dependent on the size of the cavity. Assuming a
thermal expansion coefficient for copper of $1.7\E{-5}$ per
$\degC$, a change in temperature of $\sim0.1\degC$ leads
to an expansion of $1.7\E{-6}$, equivalent to $\sim 5\kHz$ at S-band.
Typical frequency variation over the 18 hour period is shown
in Figure~\ref{fig:BPM91011_freq_decay}a. The BPM frequency showed a correlation
 with temperature (see Figure~\ref{fig:BPM91011_freq_decay}b) which is in reasonable agreement with this prediction. Despite having a similar
decay time, the variation for BPM 1 was significantly higher than
BPMs 9-11 and 2. This was found to be due to interference between 
different cavity modes as a result of the mechanical damage to the
cavity.

Variations in frequency compromise the calibration and produce an
apparent drift and a reduction in resolution if not re-calibrated. Our
studies have shown however that induced drifts and resolution deterioration
due to frequency changes are small compared to those introduced by
other effects (see Section \ref{sect:results}).

During the 18 hour period, there was a significant change in the
frequency of BPM $11x$ from $36.46 \MHz$ to $36.55 \MHz$. This change
occurred during a short access to the experimental hall, so it seems
most likely that this was caused by a mechanical disturbance of the
cavity or the associated tuner. 
During the remainder of the 18 hour running, the altered
frequency was applied where appropriate.

After we determined the frequency, we could then find the sampling
point ($t_0$ in Equation~\ref{eq:ddc_filtering}). This point is
defined as the location in the down-converted waveform where we sample
the phase and amplitude for the BPM considered. For obvious reasons,
the Gaussian filter window is centred around this sample. There is
a number of considerations when choosing the optimal sampling point
for the analysis including the maximisation of the signal-to-noise
ratio, ensuring the sampling point is away from any monopole
contamination at the beginning of the waveform and also the phase being
constant around the selected sample. We decided to pick a point $10\%$ down
from the peak of the filtered waveform (see
Figure~\ref{fig:ddc_pict}), as a trade-off between the criteria
mentioned above. We found the optimum sampling point for each BPM analysing the data containing a large range of the offsets and taking the mean of the resultant sampling point distribution.

Variation of the time difference between the digitiser trigger and the
beam arrival time would cause several systematic errors. Firstly, any
difference in frequency between  
the position and reference cavities would make the relative phase sensitive to trigger 
time variation, thus changing the $IQ$ phase. Though we had attempted to minimise this
effect by tuning both the reference and position cavities to the same
frequency, differences were still present. Secondly, the trigger time variation would change the 
relative position of the fixed sampling point with respect to the
waveform. This would cause variations in both the recorded amplitude
as well as the phase (if the phase was not stable around the chosen
sampling point) that would produce apparent changes in scale and $IQ$
phase. As the digitiser was triggered by signals synchronised with the
linac timing system, its variation was negligibly small. We determined
such variation to be less than a
fraction of the digitiser sample by measuring the timing jitter of a
reference cavity waveform rectified with a crystal detector. 

The only remaining parameter to optimise was the bandwidth of the
Gaussian filter used in the DDC algorithm ($\Delta f$ in
Equation~\ref{eq:filter_function}). To determine the optimal value
for the filter bandwidth, we calibrated the BPMs and calculated the
resolution (as defined in Section~\ref{sect:resolution}) for a range
of bandwidth values. For each BPM station, we used the bandwidth that
gave the best resolution for the remainder of the tests (see
Table~\ref{table:filter_bw}). Note that the optimal filter bandwidth 
scales with the inverse of the cavity fill time, as expected.

\begin{table}
\begin{center} 
\caption{\label{table:filter_bw} The filter bandwidth used in the DDC analysis for each set
  of BPMs.}
\vspace*{0.1cm}
\begin{tabular}{c|c} \hline 
BPM & Bandwidth ($\MHz$) \\ \hline \hline
1-2 &        1.3      \\ 
3-5 &        3.8      \\ 
9-11 &       1.3      \\ 
$Q$ cavities & 2.8      \\ 
\hline
\end{tabular}
\end{center} 
\end{table}

\subsection{Calibration}
\label{sect:calibration}

To measure the remaining calibration coefficients, $S$ and $\Phi_{IQ}$ in 
Equation~\ref{eq:pos_eq}, we induced a known change in the recorded beam position 
either through movement of the beam or the BPM. The magnitude of these
moves is restricted by the dynamic range of the BPM on the one hand and the beam
jitter and short time scale drift on the other hand. We chose typical
ranges of $\pm 500\um$.

\subsubsection{BPM calibration with mechanical movers}
\label{sect:mover_cal}
The most accurate method of calibration available was by means of the
mover system on which BPM 4 is mounted. By moving this central BPM of
the triplet and using data from the two surrounding BPMs, it is
possible to remove the beam jitter and drift from the calibration
scan in the central BPM to first order. These three BPMs are also
instrumented in the $x$ direction with an interferometer which allowed
accurate calibration of the mover system in this axis. As the accuracy
of the interferometer is $\sim 5 \nm$, the large moves of $\pm500\um$
could be established with a negligible error. 

The method used to calibrate BPM 4 with such data was
developed by the NanoBPM collaboration~\cite{ref:nanobpm}. The beam
orbit through the three BPMs (3-5) can be considered a straight line in
both $xz$ and $yz$ planes. The position in the central BPM can then be
predicted from a linear combination of the outer BPM coordinates:

\begin{equation}
  \label{eq:resol_sum}
  x_k = \delta_k + \sum_{i\neq k} (\alpha_i x_i + \beta_i y_i + \rho_i x'_i + \sigma_i y'_i)
\end{equation}

\noindent where $x_i$ and $y_i$ are the positions recorded in BPM $i$ with the
primed coordinates indicating the beam slope values and $\alpha$,
$\beta$, $\rho$, $\sigma$ and $\delta$ are constants that encode the
relative rotations, offsets, scales and $IQ$ phase differences between
the BPMs. As these should not change over short time periods, the same
constants can be used to predict the position for many
events. Consequently, this relationship can be rewritten in terms of a
matrix multiplication:

\begin{equation}
\left( \begin{array}{c} x_{k1} \\ x_{k2} \\ \vdots \\ x_{kj} \end{array} \right)
 = \left( \begin{array}{cccccc} 1 & x_{i1} & y_{i1} & x'_{i1} & y'_{i1} & \cdots \\ 
1 & x_{i2} & y_{i2} & x'_{i2} & y'_{i2} & \cdots \\ 
\vdots & \vdots & \vdots  & \vdots  & \vdots &  \\
1 & x_{ij} & y_{ij} & x'_{ij} & y'_{ij}  & \cdots
\end{array} \right) \cdot \left( \begin{array}{c} \delta_k \\ \alpha_i
 \\ \beta_i \\ \rho_i \\ \sigma_i \\ \vdots \end{array} \right),\; \mbox{or}\quad {\bf x_k
 } = {\bf M} \cdot {\bf K_k} 
\end{equation} 

\noindent where the $i$ sub-index denotes BPMs 3 and 5 and the $j$ sub-index denotes
event number. Thus, to find the constants $\alpha$, $\beta$, $\rho$, $\sigma$ and $\delta$,
the matrix of positions above must be inverted:

\begin{equation}
\label{eq:matrix_inversion}
{\bf K_k} = {\bf M^{-1}} \cdot {\bf x_k }
\end{equation}

\noindent To perform the inversion, we employed the method of Singular
Value Decomposition (SVD). This involved decomposing the (possibly
singular) $M \times N$ matrix into the product of an $M \times N$
column-orthogonal matrix, an $N \times N$ diagonal matrix with
positive or zero elements (the {\it singular values}) and the
transpose of an $N \times N$ orthogonal matrix which could then be
trivially inverted. The benefit of this method is that, as no exact
solution to Equation~\ref{eq:matrix_inversion} is possible due to its
over-constrained nature, the least squares minimisation of the
equation is found instead. Thus,
this method can be used to minimise the BPM residuals
by finding the best set of coefficients that can predict the
position in the central BPM from the surrounding BPMs.

This method can also provide predictions for the in-phase ($I$) and 
quadrature ($Q$) components of the signal. First, the positions
and slopes of the outer BPMs must be replaced with their measured
$I$s and $Q$s in Equation~\ref{eq:matrix_inversion}. Then the
deconvolution and inversion must be carried out twice: once with the
position vector (${\bf x_k}$) replaced by a vector of the measured
in-phase values of the central BPM and the second time with the
position vector being replaced by a vector of measured quadrature
values.

Assuming the BPM response is linear, the prediction from the outer 
BPMs to the central BPM is independent of any beam motion that is seen by all
BPMs. This includes both beam jitter and beam drift. However, if the
central BPM is shifted from its position relative to the outer BPMs, 
the coefficients would still predict the position at the central BPMs
original location. Thus, by comparing the measured and
predicted $I$s and $Q$s, the consequence of this move can be seen.

\begin{figure}
\begin{center}
\epsfig{file=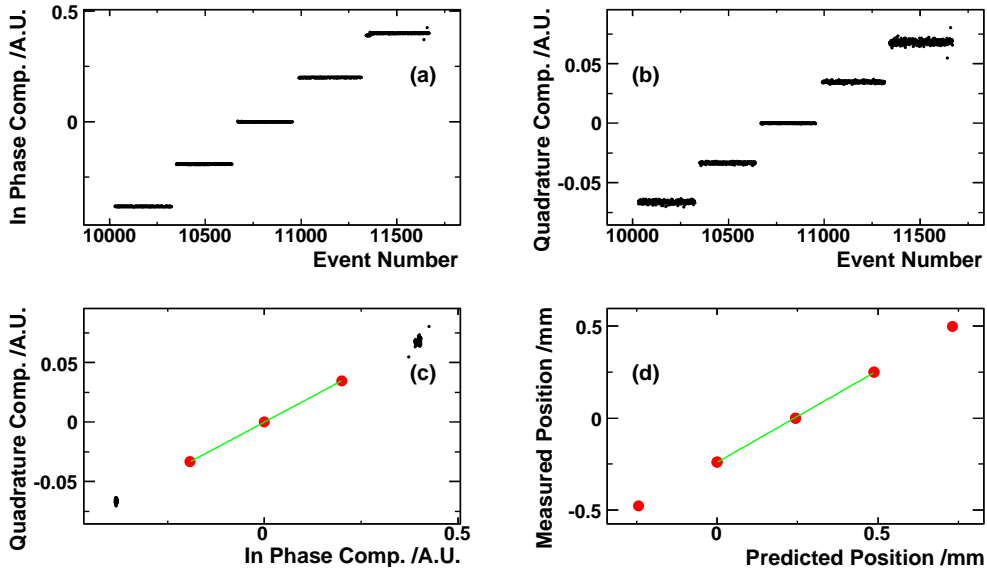,width=1.0\textwidth}
\caption{\label{fig:mover_iq} A typical mover calibration. The change
 of a) the In-Phase ($I$) and b) the Quadrature ($Q$) components against the
  event number and c) against each other. d) The predicted
  position against measured position after the calibration. }
\end{center}
\end{figure}

To perform a BPM calibration using this method, we moved the central
BPM from $-0.5\mm$ to $+0.5\mm$ in 5 steps monitoring the $x$-moves
with the interferometer. We then determined the optimum set of
coefficients to predict the $I$s and $Q$s at the central BPM
using data from the middle step in the calibration (i.e. at zero
relative offset). From these coefficients we predicted the $I$s and
$Q$s for the rest of the calibration. The mean values for $I$ and $Q$
at each mover position in the calibration were found by fitting the $I$ and
$Q$ distributions with a Gaussian function. In the $IQ$ plane these points form
a straight line (see
Figure~\ref{fig:mover_iq}). As the source of the movement was known
to be produced by a position change, this line therefore describes the
position axis and consequently, its angle of inclination
is equal to the $IQ$ phase ($\Phi_{IQ}$). The resulting predicted
position in BPM 4 can then be easily calibrated against the known mover
positions, yielding the scale $S$. In order to avoid saturation at
the extreme mover positions, we only used the central three steps of
the five step mover scan to determine the calibration constants.

\subsubsection{Corrector scan}
\label{sect:corr_scan_desc}
To calibrate the remaining BPMs, we varied the integrated field in the
correctors A32 and A33 (see Figure \ref{fig:aline_config}) from $-0.01
{\rm~kGm}$ to $+0.01 {\rm~kGm}$ in five steps. This induced a change
of between $\pm 0.5\mm$ and $\pm 1\mm$ at the downstream ESA
BPMs. We took 60-70 events per step and, for each of these, computed
the $I$s and $Q$s. Using the change in $I$ and $Q$ for the various
steps, we calculated the $IQ$ phase ($\Phi_{IQ}$) and scale ($S$)
as for the mover calibration. Again, only the central three steps
were used to avoid saturation.

\begin{figure}
\begin{center}
  \epsfig{file=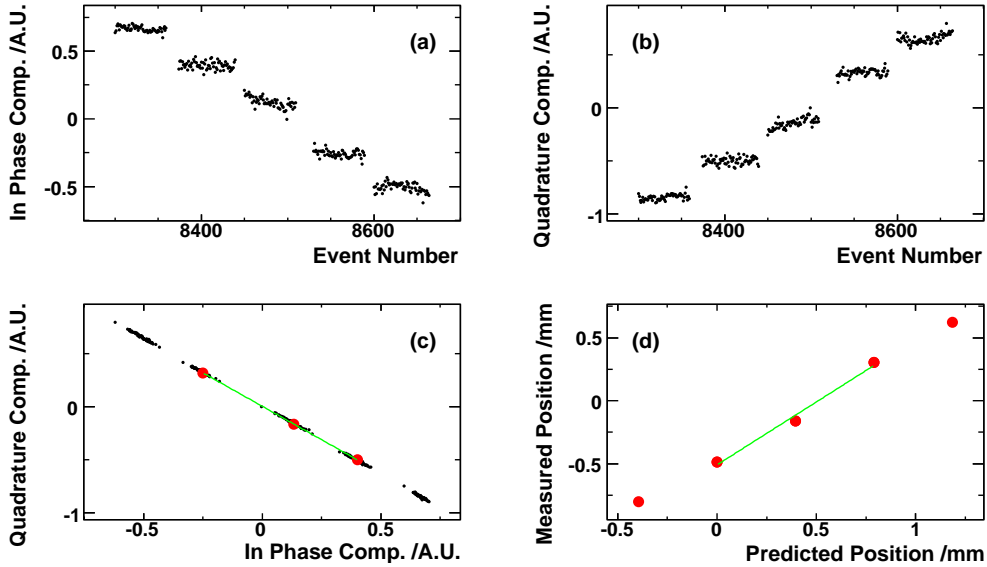,width=1.0\textwidth}
  \caption{\label{fig:corr_cal} A corrector scan calibration for BPM
    $10x$. a) The In-Phase ($I$) and b) the Quadrature ($Q$) components
    against event number, and c) against each other. d) The predicted position
    against measured position after the calibration. }
\end{center}
\end{figure}

A significant drawback of this calibration method is that the beam jitter
and drift can not be removed by using outer BPMs to predict the position of a
central BPM.
Though this has little impact on the $IQ$ phase determination, the scale 
determination is very sensitive to beam drifts, which alter the
assumed relationship between the true beam position and the magnetic
field in the corrector magnets. 
In addition, the accuracy of the
predicted beam positions was not known as the magnetic field in the
corrector magnets was not continuously monitored. The
actual changes in beam position could not therefore be determined
with sufficient accuracy and this directly 
contributed to the uncertainty on the scale. 

To estimate the improvement in scale determination when using the
movers as well as estimate the non-linearities involved, we fitted the
predicted versus measured position distribution (see
Figure~\ref{fig:mover_iq}d and Figure~\ref{fig:corr_cal}d)
with a straight line and found the maximum deviation from the
fit. This gave a residual of $\sim5\%$ for the corrector scan and
$\sim0.5\%$ for the mover scan.

In order to minimise the uncertainty on the scale from drifts during
the scan, the more accurate scale for BPM 4 as found from a mover
scan was used to correct the BPM scales found from the corrector scan:

\begin{equation}
S^\prime_i = S_i \times \frac{S_{\rm 4, mover}}{S_4}
\label{eq:scaleAdjust}
\end{equation}

\noindent where $S_i$ is the scale for BPM $i$ computed using a
corrector scan, $S_{\rm 4, mover}$ is the scale for BPM 4 found using
a mover scan and $S'_i$ is the corrected scale. The typical correction
factor was of order $15\%$. 

\section{Experimental results}
\label{sect:results}
Having described the methodology of our experiment in the previous
sections, we will here report on resolution and stability
measurements over short and long periods of operation for the separate
BPM stations as well as the complete eight-BPM system over a baseline of 32~m.

\subsection{BPM resolution}
\label{sect:resolution}

To measure the resolution of a BPM, we again used the SVD method
described in Section~\ref{sect:mover_cal} in order to minimise the
effect of any calibration inaccuracy on the measurement. The width of
the distribution of residuals however contains contributions from all
BPMs considered. Considering Equation \ref{eq:resol_sum} and making
the assumption that the contributions from both the slope and the
offset in the orthogonal axis are small we write the width as:

\begin{eqnarray}
\label{eq:sigma_svd}
\sigma_k = \sqrt{R_k^2 - \sum_{i\neq k} \alpha_i^2 R_i^2}
\end{eqnarray}

\noindent where $R_k$ is the resolution of BPM $k$.

\begin{figure}
\begin{center}
\epsfig{file=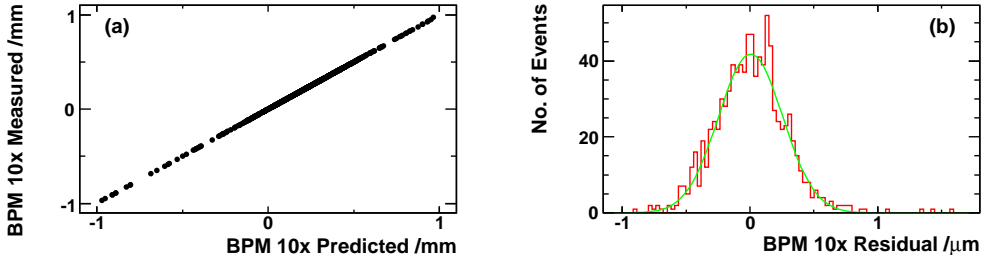,width=1.0\textwidth}
\caption{\label{fig:pred_vs_meas} a) The predicted position versus measured position
for BPM~10$x$ and b) the residual distribution for the same data.}
\end{center}
\end{figure}

With the further assumption that the predicted position in the residual
calculation is defined purely by geometry, the resolutions of all
BPMs are the same and the BPMs are equidistantly spaced, Equation
\ref{eq:sigma_svd} reduces to $R_k = \sqrt{2/3}\,\sigma_k$ for a
triplet and $R_k = \sqrt{1/2}\,\sigma_k$ for a doublet. This
correction was applied in the calculation of the individual BPM
resolution inside a BPM station. More BPMs can be included, up to
the entire set, and the resolution of the reconstructed orbit can
be determined. As some of the assumptions above are not valid in this
case, we quote the width of the distribution of residuals "as is"
without any corrections and refer to it as the ``linked" system
resolution. In the energy spectrometer, the precision of the orbit
reconstruction in the middle of the chicane directly contributes to
the energy measurement. We therefore calculated the residual for BPM~3
in the middle of the baseline versus the rest of the system. The
resolutions as recorded during a typical run are shown in
Table~\ref{table:resol_table} and an example of a residual distribution
is shown in Figure~\ref{fig:pred_vs_meas}.

\begin{table}
\begin{center} 
\caption{\label{table:resol_table} The measured and predicted
  resolutions of the BPMs in the ESA beam line. }
\vspace*{0.1cm}
\begin{tabular}{c|c|c|c|c} \hline 
BPM & Measured $x$ & Predicted $x$ & Measured $y$ & Predicted $y$ \\ 
  & ($\mu m$) & ($\mu m$) &  ($\mu m$) & ($\mu m$) \\ \hline \hline 
1, 2 & $1.1 \pm 0.1$ & $1.3\pm0.7$ & $2.2 \pm 0.3$ & $1.4\pm0.8$\\ 
3, 4, 5 & $0.53 \pm 0.05$ & $0.2\pm0.1$ & $0.46 \pm 0.02$ & $0.3\pm0.2$\\
9, 10, 11 & $0.19 \pm 0.01$ & $0.1\pm0.1$ & $0.17 \pm 0.01$ & $0.2\pm0.1$ \\ 
Linked & $0.82 \pm 0.14$ & & $1.19 \pm 0.35$ & \\
\hline
\end{tabular}
\end{center}
\end{table}

The main factors contributing to the resolution are electronic noise,
digitiser errors and mechanical motion. To estimate the
combined RF electronics and digitiser contributions,
we ran the DDC algorithm over waveforms that did not contain beam data 
(i.e. that contained only noise). Dividing the mean of these
results by the mean reference amplitude for each BPM and then
multiplying by the calibration scale, gave us the predicted resolutions
also shown in Table~\ref{table:resol_table}.

\begin{figure}
\begin{center}
\epsfig{file=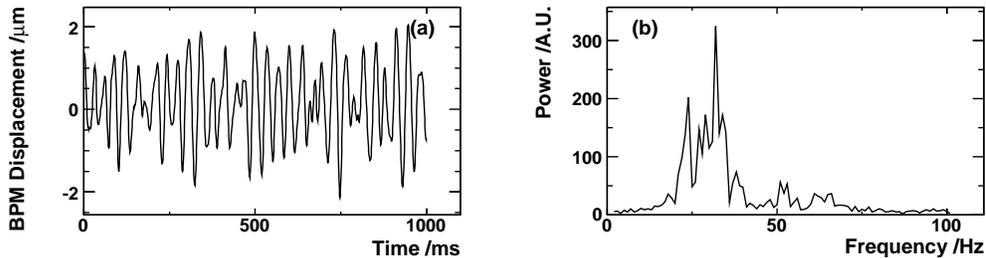,width=1.0\textwidth}
\caption{\label{fig:zygo_power} a) The displacement of BPM~4$x$ and b) the power spectrum of the vibrational motion as recorded by the interferometer running at $1\kHz$. }
\end{center}
\end{figure}

The measured resolutions for BPMs 1-2 and 9-11 are in reasonable agreement with those predicted 
indicating that the performance of these BPMs is limited by
the combined electronic and digitisation errors. 
The resolution for BPMs 3-5 is significantly worse than what is predicted
but is in reasonable agreement with the amount
of vibration recorded by the interferometer (see Table~\ref{table:zygo_motion})
thus indicating that the resolution of these BPMs is limited by
mechanical motion. In order to remove this effect, we included the
interferometer data in the matrix analysis as additional variables in
Equation~\ref{eq:resol_sum}. This addition improved the
resolution measurement from $0.53\um$ to $0.45\um$ (taking into account the geometric factor), 
which is equivalent to the removal
of $0.35\um$ of mechanical vibration. This is only half of the non-rigid body motion
recorded by the interferometer. High rate ($1\kHz$) interferometer data indicated that the
horizontal vibrational power spectrum peaked between 
$20-40\Hz$ (see Figure~\ref{fig:zygo_power}). The latency
between interferometer and BPM data arrival acquisition time 
is of the same order so not all of the mechanical
motion could be removed. We plan to remedy this deficiency in future 
runs.

\subsection{Calibration stability}
\label{sect:calibration_stab}

We took several corrector and mover calibration scans over the course
of the dedicated 18 hours of operation in order to study the stability
of the calibration coefficients. As the parameters of the system vary
over time, these coefficients maintain their validity only for a
limited period, after which a recalibration is
necessary. Understanding of these effects is important for long term
stable operation of the BPM system in the spectrometer.

\begin{figure}
\begin{center}
\epsfig{file=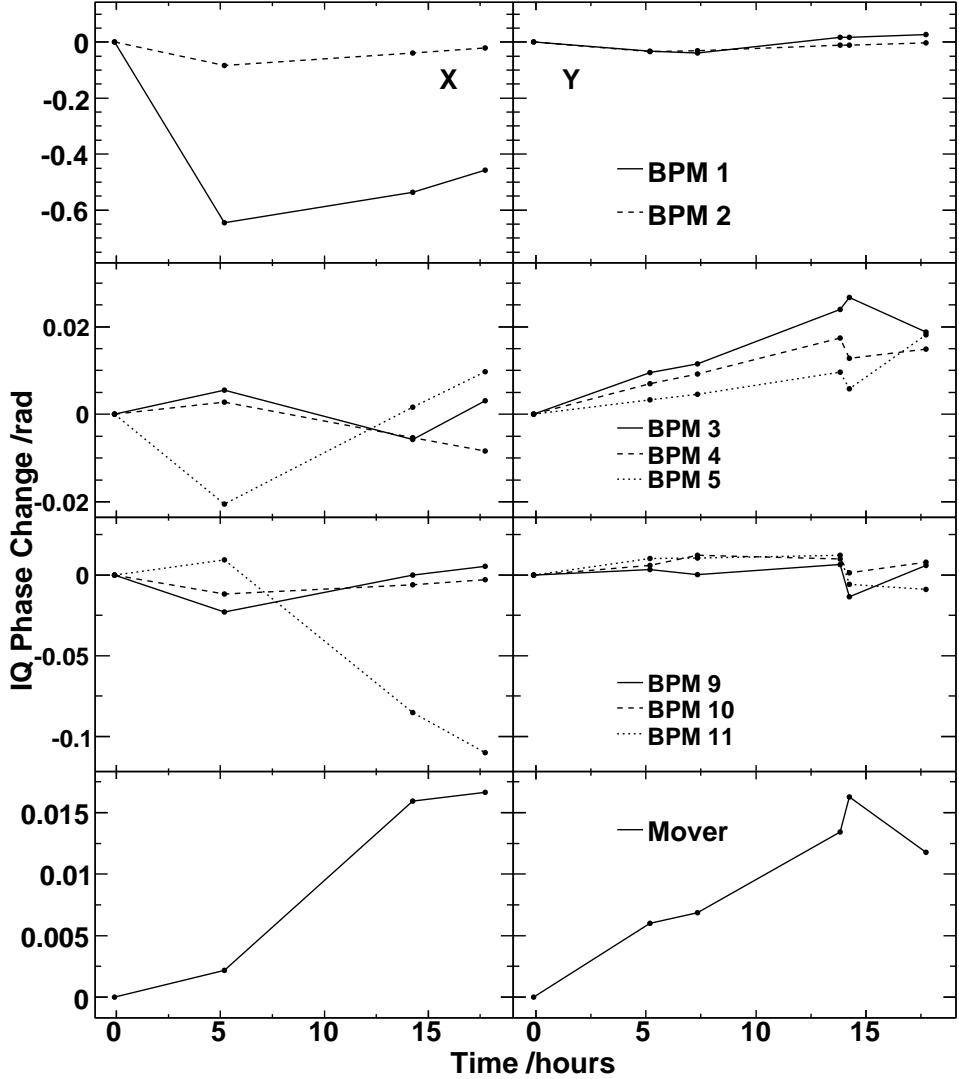,width=\textwidth}
\caption{\label{fig:full_cal_stab_iq} Variation of IQ rotation 
observed over 18 hours of operation.}
\end{center}
\end{figure}

\begin{figure}
\begin{center}
\epsfig{file=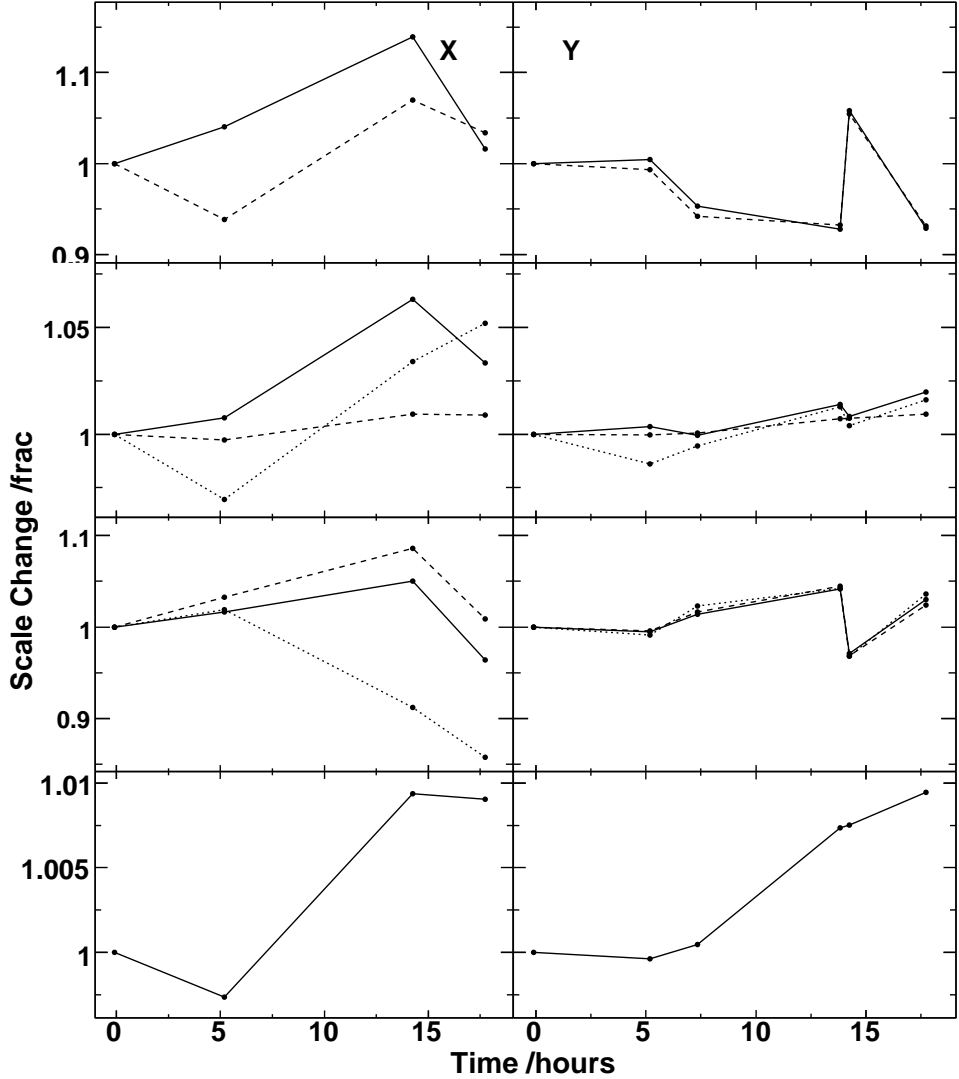,width=\textwidth}
\caption{\label{fig:full_cal_stab_scale} Variation of scale 
observed over 18 hours of operation.}
\end{center}
\end{figure}

The variation of the calibration constants over the 18 hour running period is shown in
Figures~\ref{fig:full_cal_stab_iq} and~\ref{fig:full_cal_stab_scale}.
The $IQ$ phase variation is small in all the BPMs except
for $1x$ and $11x$. The large phase change in BPM $1x$ was due to a small change in the digitiser
trigger time relative to the beam arrival time. Other BPMs are less
less sensitive to trigger fluctuations.
As mentioned in Section~\ref{sect:cavity_specs} however, BPM $1x$ had significant
cross-coupling from the vertical mode which resulted in the measured phase being
time-dependent. This sensitivity produced a large change in $IQ$ phase during the first
calibration run and so this calibration for BPM $1x$ was not used.
The large changes in both $IQ$ phase and scale for BPM $11x$ are the
result of the perturbation already discussed in
Section~\ref{sect:freq_decay}. Though a new frequency
was used for this section of the run, there is still some residual change. Consequently,
we used a separate calibration (frequencies, $IQ$ phase and scale) for 
this BPM from the time of the perturbation onwards.

Apart from those exceptions noted above, the $IQ$ phase for most BPMs
indicated a drift of about $20\mrad$ with no significant difference
between the mover and corrector calibration. 
The scale variation for the BPMs calibrated using the correctors was large
despite the corrections of Equation~\ref{eq:scaleAdjust}.
In the $x$ direction, these variations
were consistent with the statistical error (about $\pm4\%$). In the 
$y$ direction
the variation was larger than could be accounted for from beam jitter
alone. The scale variations were anti-correlated for the first and third stations
which suggested there was a angular variation during the corrector
scans with the pivot point somewhere between these BPM stations. 
Since only BPM 4 was equipped with a mover system, it was
only possible to remove offset drifts, not angle drifts, using 
Equation~\ref{eq:scaleAdjust}. To measure the slope drift during a
calibration run,  
we used the measured $y$ positions
of BPMs 9-11 and BPMs 1 and 2 to calculate the overall slope of the beam and found changes
of up to $\sim3\urad$. Over the $\sim20\m$ of beam line between the central BPM station (3-5)
and the outer BPM stations, this leads to a change of $\sim 60\um$ during a calibration
step, introducing a change in scale of $\sim10-15\%$. This is in good agreement with the
$y$ scale variation observed.

\begin{figure}
\begin{center}
\epsfig{file=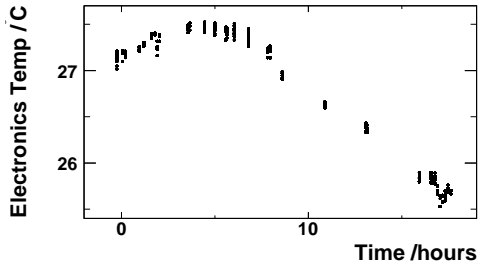,width=0.5\textwidth}
\caption{\label{fig:esa_temps} The temperature at the electronics racks in the 
experimental hall during the 18 hour run.}
\end{center}
\end{figure}

The $\mathcal{O}(1\%)$ scale variations observed in the mover
calibrations appear to be correlated to the temperature of the
electronics racks (see Figure~\ref{fig:esa_temps}). 
To further test whether environmental effects were behind the gain
variations in the electronics, we applied a constant CW tone 
to the electronics 
for both the dipole and reference cavities and tracked drifts in relative
amplitudes over the course of several hours. A variation of similar magnitude was found from these tests.

As the variation in $IQ$ phase was small and the large scale changes
seen in the corrector calibrations seem to be caused by the beam
drifts rather than electronics drifts, we averaged the coefficients
obtained from the separate calibrations and used the mean values 
for the entire 18 hour
run. The only exceptions to this were BPM $1x$ 
in which the
first calibration was removed from the average and BPM $11x$ where two calibrations
were used, one computed from calibrations before the 
mechanical perturbation and the other from after it.

\subsection{Stability of BPM offsets and resolution}
\label{sect:stability}
\begin{figure}
\begin{center}
\epsfig{file=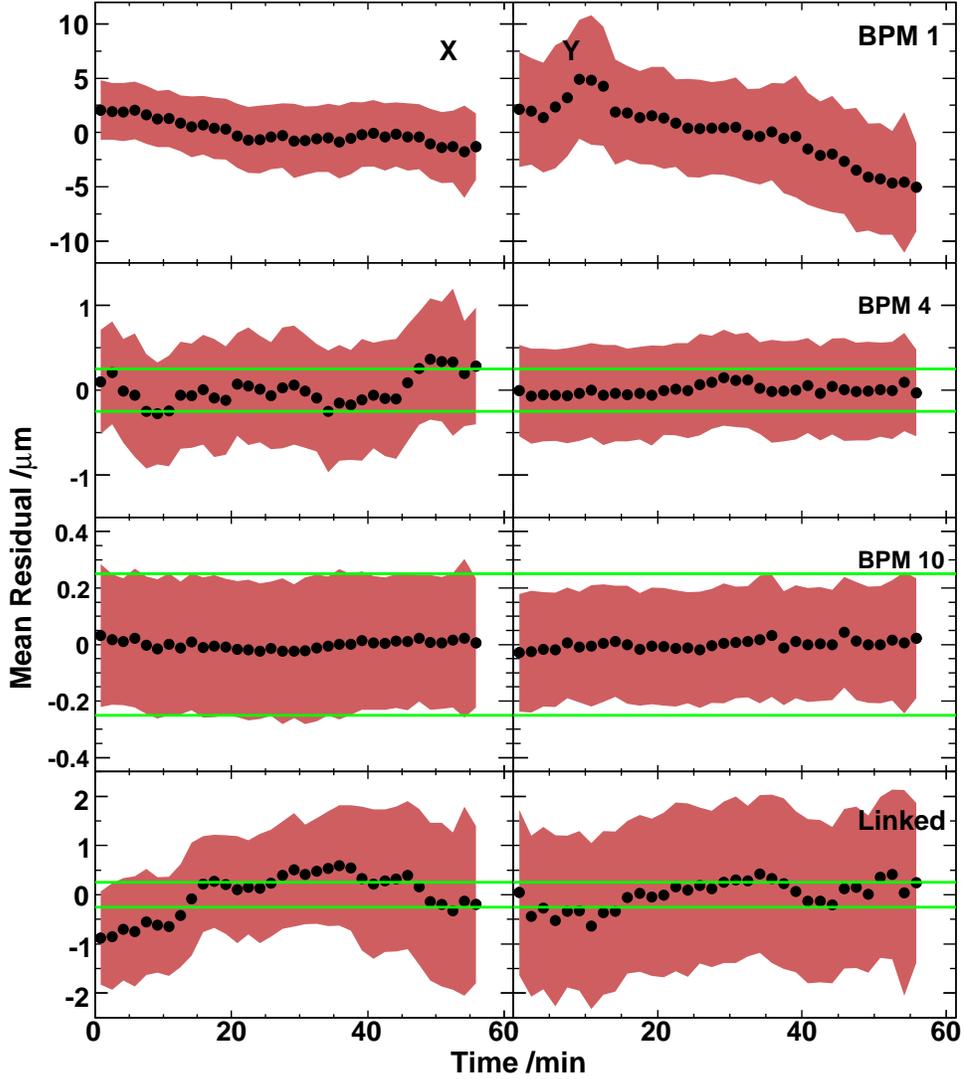,width=\textwidth}
\caption{\label{fig:full_stab_short} The RMS and mean residual measured in
different BPM stations and for the whole system over 1 hour
of operation. Each point corresponds to 1000
events.}
\end{center}
\end{figure}

\begin{figure}
\begin{center}
\epsfig{file=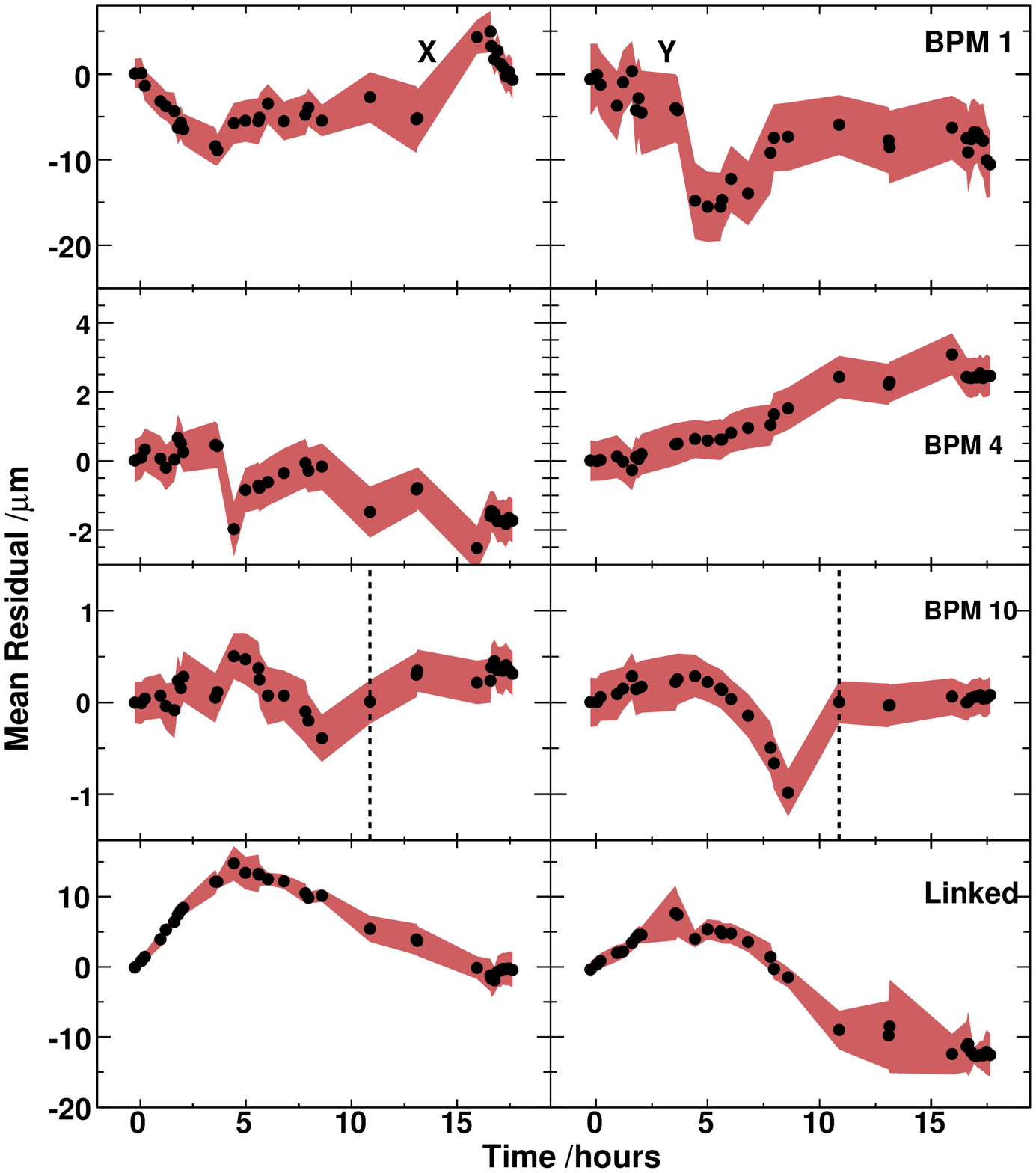,width=\textwidth}
\caption{\label{fig:full_stab_long} The RMS and mean residual measured in
different BPM stations and for the whole system over 18 hours
of operation. Each point corresponds to 1000
events. The dotted line indicates where the SVD coefficients were recomputed
due to the perturbation of BPM $11x$.}
\end{center}
\end{figure}

We investigated the stability of the BPM system over both short and
long periods of operation. In both cases the calibration was
refined using the SVD over the first 1000-event block of the data.
The 1 hour data shown in
Figure~\ref{fig:full_stab_short} is a zoom into the last hour of the 18 hour
run (Figure~\ref{fig:full_stab_long}) after recomputing the SVD coefficients.  We
had to select the periods of stable data-taking excluding the data
taken during the machine tuning, big energy jumps due to klystron
failure etc. as well as the calibration data. In addition, due to the perturbation
of BPM $11x$, the SVD coefficients for BPMs 9-11 were recomputed.

A 100~nm stability of BPMs~9-11, 500~nm of BPMs~3-5 and micron
stability of the full system are observed over one hour of
operation. The limiting factor for the system's stability is the
stability of BPMs~1 and 2, the electronics for which was exposed to
large temperature variations.

\begin{figure}
\begin{center}
\epsfig{file=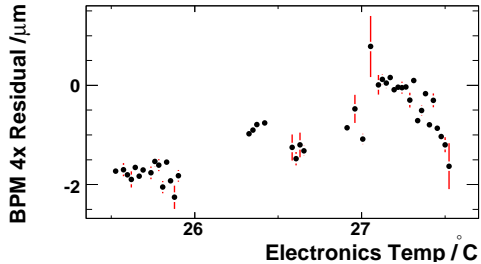,width=0.5\textwidth}
\caption{\label{fig:res_vs_temp} Residual of BPM $4x$ against electronics temperature.
Events within $\pm25\um$ of the mean position over 18 hours of operation are shown.}
\end{center}
\end{figure}

For all the BPM stations, the RMS of the residual does not show any significant change
with time over the course of the 18 hour run. In contrast, the mean
residual experiences large variations, especially for BPMs~1 and 2, and
therefore for the linked system as well. We considered the most likely source of these
drifts to be gain variations in the electronics. 
This is supported by Figure
\ref{fig:res_vs_temp} where we show the BPM~4 mean residual versus
electronics temperature for events in which the beam position in 
BPM~4 was within $\pm25\um$ from the mean position over the course of the 18 hour run. Though
there is a change in behaviour at temperatures above $27^{\circ}$C, below this
value a significant correlation is seen.

\begin{figure}
\begin{center}
\epsfig{file=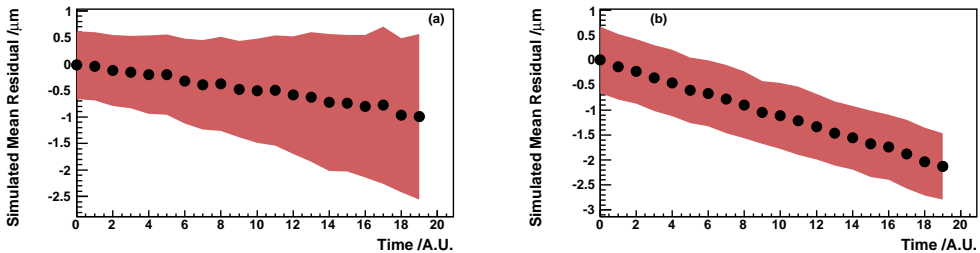,width=1.0\textwidth}
\caption{\label{fig:sim_results} Mean residuals from a simulated BPM triplet experiencing
a $1\%$ scale change in a) the central BPM, b) all three BPMs.}
\end{center}
\end{figure}

To estimate the behaviour of a triplet of BPMs when subject to changes
in scale, we simulated three BPMs with offsets typical in our system
and a beam orbit experiencing typical beam jitter and drifts. A scale
simulated scale drift of $1\%$ was implemented, similar in size to the 
effects seen in the experiment.
We looked
at two scenarios: only the central BPM's scale drifting only (see
Figure~\ref{fig:sim_results}a) and all three BPM scales drifting by
the same amount and in the same direction (see
Figure~\ref{fig:sim_results}b). In the first case, both the RMS value
and the mean of the apparent beam residual increased gradually, while 
in the second simulation the RMS
value remained almost the same, while the offset increased 
gradually. The second case is consistent with
the effects observed in our system: the
physical offsets of the BPMs with respect to each other reconstructed
by temperature-dependent gains, 
resulting in observed changes in the reconstructed offsets.

We tried improving our stability studies for the central BPM station
including the interferometer data into SVD computations. Unfortunately
this effort failed as the drifts over long periods of time observed in
the interferometer seem to be caused by the thermal expansion of the 
supporting aluminium table and the change of the refractive
index of the air rather than the actual mechanical motion.

\begin{figure}
\begin{center}
\epsfig{file=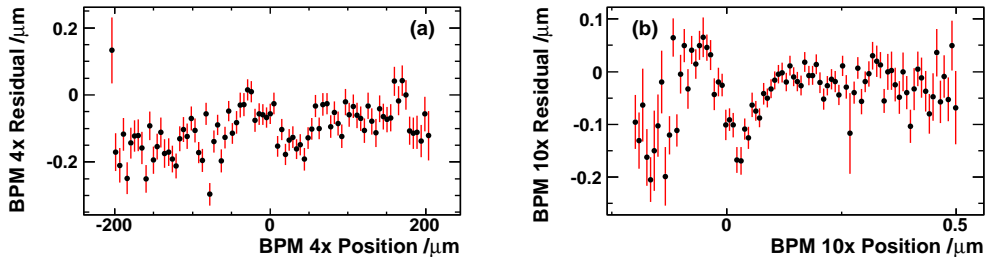,width=1.0\textwidth}
\caption{\label{fig:corr2} a) The residual between the BPM~4$x$
  position measured and predicted with BPMs 3 and 5 over the course of
  1 hour against BPM~4$x$ position. b) The residual between the BPM
  10$x$ position measured and predicted from BPMs 9 and 11 over the
  course of 1 hour against BPM~10$x$ position.}
\end{center}
\end{figure}

We noticed that the residual shows some correlation with the position
for low amplitude signals when the beam is close to the cavity centre
(see Figure~\ref{fig:corr2}),
which can be caused by contributions from other cavity modes. This
effect is small, but may contribute to the observed main residual
variation.

\begin{figure}
\begin{center}
\epsfig{file=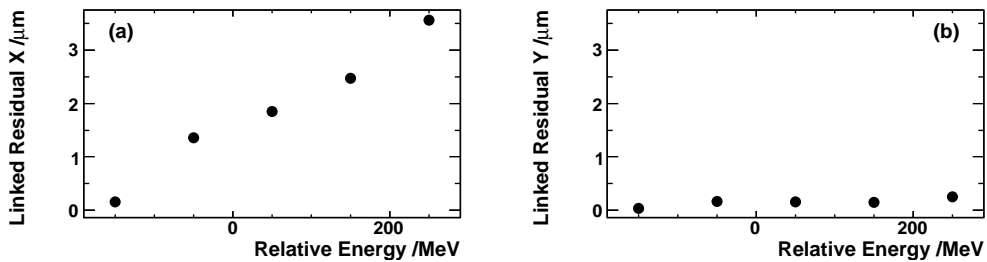,width=1.0\textwidth}
\caption{\label{fig:energy_dep} The residual distribution of the linked system at various
beam energies.}
\end{center}
\end{figure}

The drifts seen in the linked system seemed to be dominated by the
scale changes. However, over the long baseline, it is possible that
variations in energy combined with the Earth's Magnetic
Field produce a similar effect.
To check this the energy of the beam was changed from
$-150\MeV$ to $+250\MeV$ in five steps. This scan was tracked by the
system as a change of the linked system mean residual (Figure
\ref{fig:energy_dep}).

Using a set of flux gate magnetometers, we measured the stray magnetic fields present
around the beam line. Using the measured fields, a change in position
of $3.1 \um$ in $x$ and $0.1\um$ in  
$y$ is predicted for the $400\MeV$ energy change. We measured a change of $\sim 3.5\um$ in 
$x$ with no significant change 
in $y$ observed  (see Figure~\ref{fig:energy_dep}) which is in good
agreement with the prediction. The energy variation over the 18 hour
run was $\sim\pm100\MeV$ and so the drifts seen
over this time period were not consistent with a change in energy. 

\section{Conclusions}

We have successfully commissioned eight cavity BPMs of differing designs and 
properties divided into three BPM stations in the End Station A beamline at SLAC. 
The first BPM station consisted of two rectangular
cavity BPMs originally designed for use in the A-line. They demonstrated resolutions in $x$ and $y$
of $1.1\um$ and $2.2\um$ respectively and were stable to $\pm5\um$ over one hour while drifting by $\pm10\um$ over 18 hours of operation.
The resolution was limited by digitiser errors and noise. The drifts can be explained by the changes of electronics gains.

The second BPM station consisted of three prototype ILC linac cylindrical BPMs with the central 
BPM mounted on a dual-axis mover. 
These demonstrated resolutions of $0.53\um$ in $x$ and $0.46\um$ in $y$. Mechanical motion
was found to dominate the resolution. 
These BPMs were stable to $\pm0.25\um$ over a period of 1
hour and $\pm1\um$ over a period of 18 hours. The stability was influenced by
low amplitude effects and mechanical vibration on short time scales and scale changes of $\sim1\%$ magnitude over long time scales.

The final BPM station in the beamline consisted of three rectangular cavity BPMs, originally designed for use in the SLAC linac. They demonstrated resolutions of $0.19\um$
in $x$ and $0.17\um$ in $y$ with a stability of $\pm50\nm$ over the one hour period.
Long term stability of this station could only be measured over 10 hours due to 
a perturbation that altered the calibration of BPM $11x$.
However, over these 10 hours, the triplet achieved a stability of $\pm500 \nm$ in $x$
and $\pm750\nm$ in $y$. As with BPMs 3-5, we think the stability was limited by low amplitude effects and scale drifts, but there was no information on the mechanical motion available for this triplet.

When combining all the BPM stations to measure the precision of the orbit reconstruction over the whole baseline, a resolution of $0.82\um$ in $x$ and $1.19\um$ in $y$ was achieved.
The system was stable at the micron level over the course of one hour. The long term stability was affected by relative scale drifts across all the BPMs therefore drifts of the order of $\pm10\um$ were observed over 18 hours of operation.

In order to improve the system we had in 2006 and be able to perform full
tests of a magnetic spectrometer prototype, we have added several upgrades to
the beamline and BPM systems.
Four steel core dipole magnets have been installed to form the magnetic
chicane with BPMs 3 and 5 now measuring the incoming beam position, BPMs 9-11 measuring
the outgoing beam position and BPM 4 having been moved to measure the beam 
position at the mid-chicane location. 
To improve the stability of the BPMs, we have added a 
periodic sine wave calibration tone. This is applied through the BPM electronics at a 
rate of $\sim0.1\Hz$ and allows continuous gain and phase monitoring. This should allow
us to track calibration changes online and correct for any variation.
The interferometer has been
extended to measure mechanical drifts between the new incoming BPM station (BPMs 3 and 5)
and the mid-chicane BPM. Helmholtz coils have been installed to quickly dither the beam
and therefore reduce the dependence of the calibration procedure on beam drifts.
We have commissioned these upgrades during 2007 and are planning to continue in 2008, so we will report on the results in our next publication.

\section{Acknowledgements}

We wish to thank the SLAC accelerator and End Station A operations and support staff.
Our work has been supported by the U.S. Department of Energy under contract
numbers DE-AC02-76SF00515 (SLAC) and DE-FG02-03ER41279 (UC Berkeley), 
by the EC under FP6 ``Research
Infrastructure Action - Structuring the European Research Area'' EuroTev DS
Project under contract number RIDS-011899 and by the Science and Technology Facilities
Council (STFC).


\end{document}